\documentclass[twocolumn,reprint,floatfix,prd]{revtex4}%

\usepackage{amssymb,hyperref,amsmath}
\usepackage[dvips]{color}
\usepackage[dvips]{graphicx}
\usepackage{epsfig}
\DeclareMathOperator{\Tr}{Tr}
\preprint{TCDMATH-12-ZZZ}

\usepackage{latexsym} 

\newcommand {\onehalf}{\frac{1}{2}}

\begin{document}

\bibliographystyle{apsrev}

\title{A comparison of analysis techniques for extracting
    resonance parameters from lattice Monte Carlo data}

\author{Pietro Giudice}
\affiliation{Department of Physics, College of Science, Swansea University,
  \sl Singleton Park, Swansea SA2 8PP, UK.}
\author{Darran McManus and Michael Peardon}
\affiliation{School of Mathematics, Trinity College, Dublin 2, Ireland.}

\date{\today}


\begin{abstract}
Different methods for extracting resonance parameters from Euclidean lattice 
field theory are tested.  Monte Carlo simulations of the $O(4)$ non-linear 
sigma model are used to generate energy spectra in a range of different volumes
both below and above the inelastic threshold.  The applicability of the 
analysis methods in the elastic region is compared. Problems which arise in 
the inelastic region are also emphasised.
\end{abstract}

\maketitle

\section{Introduction}
A lattice regularisation of the path integral over quantum fields provides a
useful framework for investigating non-perturbative properties of the theory.
If the theory is discretised on a lattice of finite extent, the path integral
becomes a finite-dimensional integration problem and if Euclidean metric is
used, the non-negative weight of each field configuration can be used as a 
sampling measure for efficient Monte Carlo calculations. This route is widely
used in most numerical lattice calculations. 

A better understanding of scattering in the strong interaction is crucial if a
qualitative understanding of the internal structure of states of QCD is to be 
uncovered. In particular, if states with intrinsic excitations of the quarks 
and gluons are to be studied, methods for characterising the properties of 
resonances are needed since these excited states are above thresholds
for decays via the strong interaction. 

Formulating the theory in Euclidean spacetime has a drawback. Direct access to
information about dynamical processes such as scattering and the decay of an
unstable state is obscured~\cite{Maiani:1990ca}. 
Extracting information about scattering from a two-point correlation function 
at large Euclidean-time separations is not straightforward and can not be done
directly. A theoretical framework that enables computation of elastic 
scattering properties from Euclidean field theory was developed by L\"uscher. 
In L\"uscher's method, if accurate data on the discrete spectrum of
states in a finite volume can be obtained, preferably for a range of different
volumes, then scattering properties can be deduced indirectly. More recently,
new analysis paths have been suggested that take a more intuitive, direct 
approach to analysing the same spectrum data~\cite{Bernard:2008ax}: 
levels are used to estimate the density of
states in energy ranges and presented in a histogram. Once care is taken to 
subtract the background, resonance features can emerge, usually resembling the 
Breit-Wigner distribution. The validity of the histogram method follows from 
L\"uscher's analysis. 

Ref~\cite{Meissner:2010rq} proposes determining the
parameters of a resonance by fitting the correlator directly; this has the 
evident advantage that a single simulation in one volume is needed.

Over recent years, many more calculations of scattering in lattice QCD have 
been made and new measurement techniques have improved the prospects of 
performing precise computations of scattering in QCD. 
The main focus of these determinations~\cite{McNeile:2002fh,Feng:2010es,
Aoki:2011yj,Lang:2011mn} has been to use the $rho$ meson as a test case, and 
to investigate P-wave $\pi-\pi$ scattering close to this resonance. 
In spite of recent progress, the subject is still regarded as in its infancy 
primarily due to the difficulties in making suitable Monte Carlo measurements
from QCD with light dynamical quark fields of correlation functions with
more than one particle in the creation operators. 
Some progress in the operator creation methodology 
\cite{Peardon:2009gh,Morningstar:2011ka} has been made recently making 
precision Monte Carlo simulations in QCD feasible. For a recent review, see 
Ref~\cite{Bulava:2011uk}.  Motived by both the technical 
challenges and recent progress, a search for the best analysis path to take 
seems very timely.

Since QCD has pions that are much lighter than the intrinsic scale for internal hadronic excitations, the problem of studying resonances above inelastic 
thresholds needs to be addressed in a robust way. In the inelastic region, 
L\"uscher's formulation can not be applied since it relies on linking data from
quantum field theory to quantum mechanics, where inelastic scattering does not
feature. For promising generalisations of Luscher's method to
multi-channel scattering see, e.g., 
Ref~\cite{Hansen:2012tf} and Ref~\cite{Briceno:2012yi}.
The more intuitive ideas of interpreting spectrum data from different
volumes might give a new direction for determining resonance widths in 
the inelastic region. 

This paper aims to compare proposed methods with Monte Carlo data to determine
whether they agree and whether the precision obtained from different methods is comparable. We generate data in the $O(4)$ sigma model on the lattice, where
it is straightforward to choose parameters of the model to ensure resonances
emerge in lattice data. Simulations in the inelastic region are also performed 
to see if the histogram method can help to infer something about the width of
a high-lying resonace. A number of technical issues in the construction of the
appropriate lattice measurements and challenges in analysing lattice data are
observed and addressed. Preliminary progress from this work is presented in 
Refs~\cite{Giudice:2010zz} and~\cite{Giudice:2010ch}.

The paper is organised as follows. Section~\ref{sectheory} discusses the
theoretical issues surrounding resonances on the lattice, as well as
giving a heuristic derivation of both methods. In Section~\ref{themodel}, 
the model used is discussed, including the relation between the Lagrangian 
fields and the particle spectrum. The Monte Carlo simulations and the 
applications of the two methods are discussed in Section~\ref{secmontecarlo}, 
along with the results obtain from both methods.  A third method is briefly 
discussed. Finally we draw some conclusions in Section~\ref{secconclusion}.

\section{Theoretical background}
\label{sectheory}
Before describing our Monte Carlo simulations, we review a few important 
aspects of the theoretical background to studying scattering in Euclidean 
lattice field theory. 

\subsection{Two non-interacting particles in a box}

We discuss the dispersion relation of two identical non-interacting bosons
when they are in a box of volume $V=\prod_{i=1}^3 L_i$, as function of
the dimension of the box, both in the continum and in the lattice case.

In the continuum, the particles of mass $m_\pi$ characterised by a 
relative momentum $\vec{p}$, have a total energy $E$ given by
\begin{equation}
E=2 \sqrt{m_\pi^2+\vec{p}^2} \ ,
\label{eqEcont}
\end{equation}
where due to the finite volume, the momenta $p_i$ are given by
$p_i=\frac{2 \pi}{L_i} n_i$, with $n_i \in Z$.
On the lattice, the correct expression for the simplest discretisation of 
the free scalar field is 
\begin{equation}
E=4 \sinh^{-1}\left[{\onehalf \sqrt{m_{\pi,r}^2+\tilde{p}^2}}\right] \ ,
\label{eqElatt}
\end{equation}
where $\tilde{p}_i=2 \sin{\frac{\pi}{L_i}n_i}$ and $m_{\pi,r}$ is the 
``subtracted'' mass of the pion; the reason of this name will be clarified 
in the context of interacting particles in Sec~\ref{subsechisres} where 
Eq.~\ref{eqElatt} can also be used.
%
\begin{figure}[htb]
\begin{center}
  \includegraphics[width=0.36\textwidth,angle=-90]
{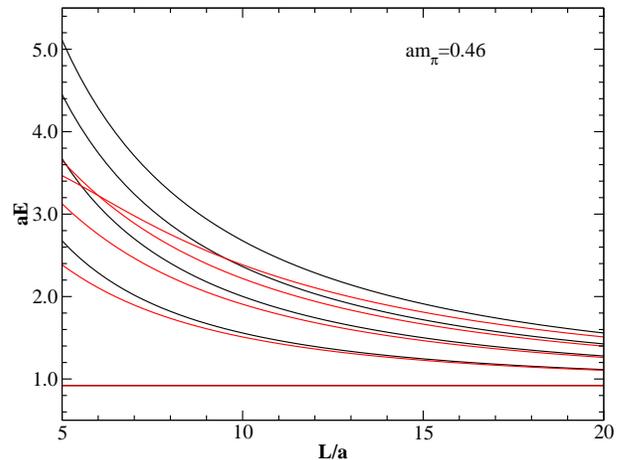}
\caption{The total energy $E$ for four different levels in the continuum 
(black lines) and in the lattice (red lines) case versus $L$.}
\label{spectr_cont_latt}
\end{center}
\end{figure}
%
We will focus in the following on the case of a cubic lattice, 
characterised by a single side length $L$; it is valuable to remember that,
in a cubic box if $n^2=\sum_{i=1}^3 n_i^2$ is fixed, degenerate
energy levels for different values of $n_i$ can appear.
Sometimes we will refer to a specific level writing the three component
vector as $(n_x,n_y,n_z)$.

It is clear that the space-time discretization can have a strong effect
in particular when the volume is small, i.e. large momentum, and $m_\pi$ is big.
In Figure~\ref{spectr_cont_latt} we show a plot of the two formulas where
it is evident that, for small volume ($L \lesssim 15$) and a 
mass $am_\pi=0.46$, the two spectra are very different; therefore we cannot 
use the continuum formula to describe our Monte Carlo results.

Note that in a general theory, such as QCD, where an expression like 
Eq.~\ref{eqElatt} is not available, we need to determine 
the non-zero-momentum single-particle energy levels 
numerically and then, to determine the 
two-particle energy spectrum, we simply multiply the results by a factor of two.

\subsection{Avoided level crossing}
\begin{figure}[ht]
  \footnotesize
  \hspace{-2mm}
  \includegraphics[width=0.36\textwidth,angle=-90]
{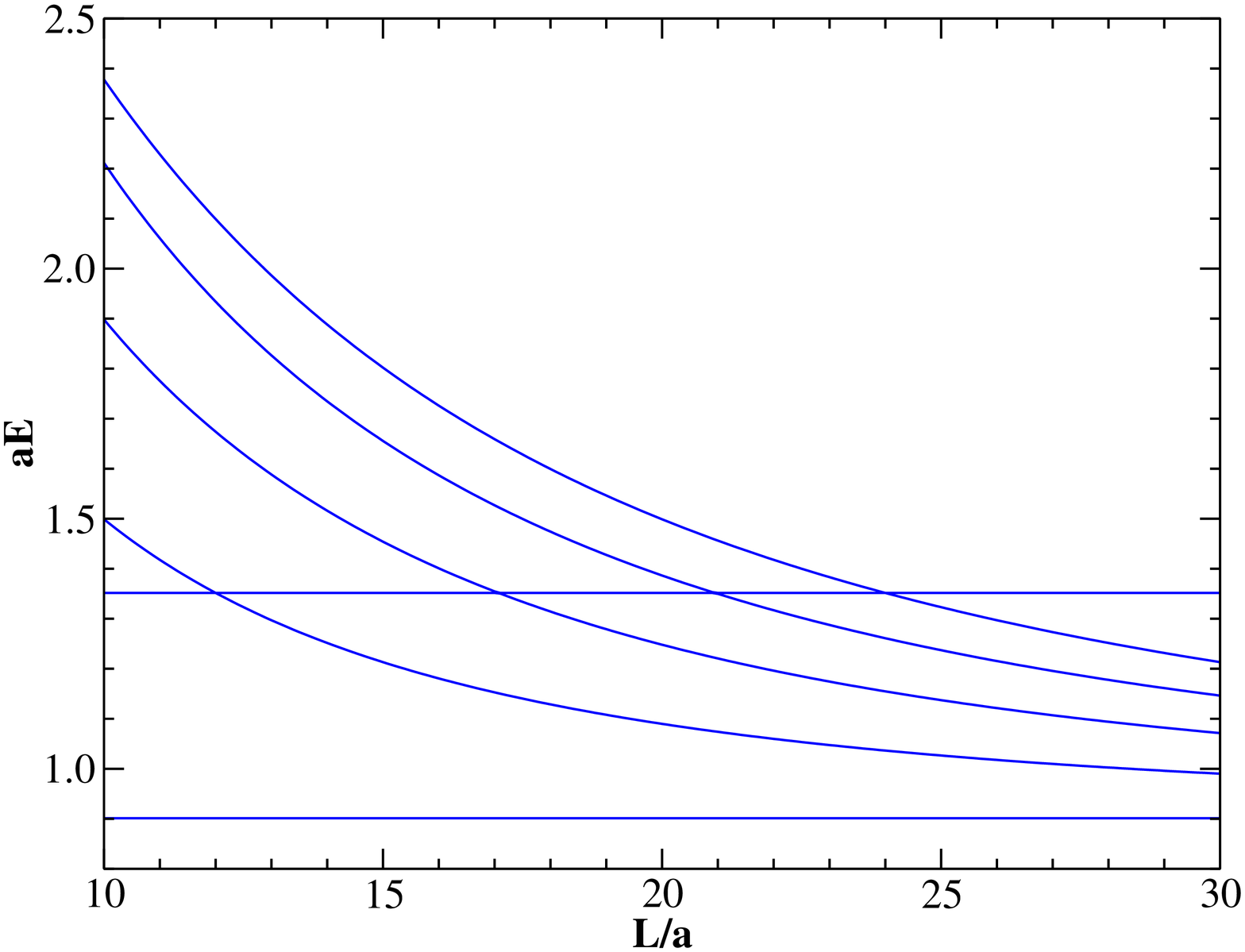}
\hspace{2mm}
  \includegraphics[width=0.36\textwidth,angle=-90]
{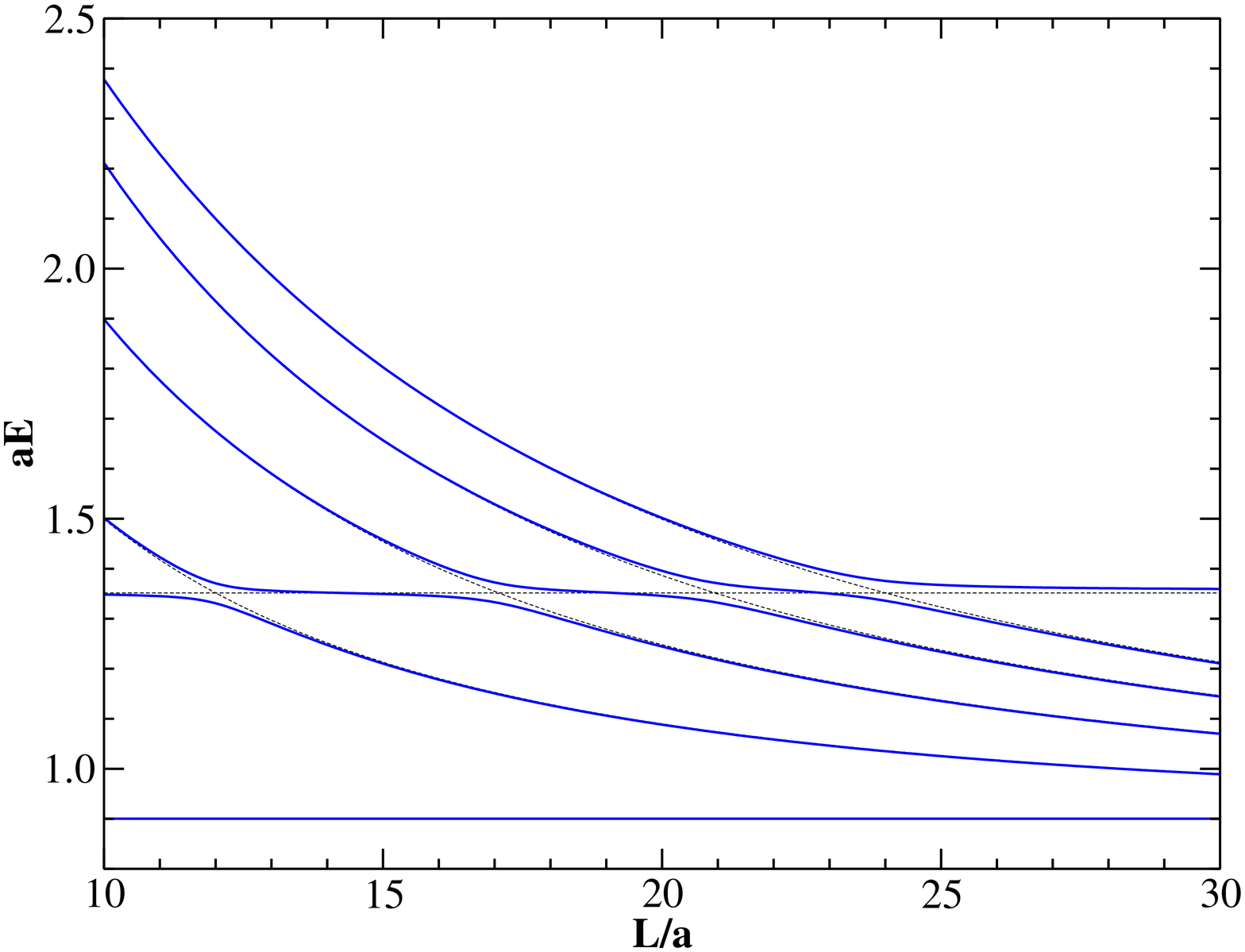}
\caption{(Top) The spectrum of a system of two non-interacting particles
of mass $am_\pi=0.4544$ worked out using Eq.~\protect\ref{eqElatt}; the
horizontal line describes the particle $\sigma$ at rest with mass
$am_\sigma=1.3517$.
With these parameters the intersection between $\sigma$ and the two-particle
level $n^2=1$, i.e. $(1,0,0)$, is set at $L=12$. 
  (Bottom) Avoided level crossings where on the (Top) there were 
intersections between $\sigma$ and $2\pi$.}
\label{spectrum_tot}
\end{figure}

Let us introduce another particle $\sigma$ in the box (at the moment,
not interacting)
with mass $m_\sigma$; we are interested in studying the 
\emph{elastic} scattering between the $\pi$ particles, therefore we 
impose the constraint $2 m_\pi < m_\sigma < 4 m_\pi$.
In Figure~\ref{spectrum_tot} (Top)
the $\sigma$ energy level is the horizontal line that
intersects the two-particle levels at various system sizes $L$.

In Minkowski space if we introduce a three point interaction $\sigma\pi\pi$ 
between the fields, the $\sigma$ will become an unstable particle, 
a resonance, and decay into two $\pi$ particles.  
In Euclidean space and in a finite volume the scenario is different. 
First of all, the finite volume will prevent the $\sigma$ from being
a resonance. This can be seen from two complimentary points of view. 
First of all, resonances appear as poles on the second Riemann sheet 
of the S-matrix. This second sheet is found by continuing through the
multiparticle branch cut. However, in a finite volume the branch 
dissolves into a series of poles and hence this second Riemann sheet is 
lost, so the $\sigma$ may only appear as a pole on the physical sheet.
Secondly, in a finite volume only certain discrete values of momenta are 
allowed. Conservation of momenta may require the two particles produced 
by the decay of the sigma to take on momenta outside these values, hence 
rendering the $\sigma$ stable. However because of the interaction, the 
energy eigenstates are a mixture of this stable $\sigma$ and the $2\pi$ 
Fock-states. One might attempt to avoid the fact that the $\sigma$ has
become a stable state, by measuring on the lattice some appropriate 
$n$-point function which contains infinite-volume scattering information, 
such as the phase shift $\delta(p)$. However the Maiani-Testa 
theorem, Ref~\cite{Maiani:1990ca}, forbids this, as the Euclidean 
$n$-point functions lack the non-trivial complex phase which 
would directly characterise $\delta(p)$.
Instead we turn to the effect that the mixing of the $\sigma$ has on 
the finite volume energy levels, which does contain information on its 
behaviour as a resonance in infinite volume: the most obvious feature 
is the avoided level crossings (ALCs), as shown in the lower panel of 
Figure~\ref{spectrum_tot}.

\subsubsection{ALC simple model}

There is a very simple model that can be use both as a method to plot
a spectrum where the ALCs are present but also as a model to test 
numerical methods for extracting the resonance 
parameters. The model is based on this correlation matrix:
\begin{equation}
{\mathcal C} = \left(
\begin{array}{ccc}
m_\sigma   & \lambda    & \lambda \\
\lambda   & E_{(0,0,0)} & 0       \\
\lambda   & 0          & E_{(1,0,0)}
\end{array}
\right),
\end{equation}
where $E_{(i,j,k)}$ are given by Eq.~\ref{eqElatt}, here $i^2+j^2+k^2=n^2$.
The diagonal terms correspond to the three lowest energy states of a 
theory where the $\sigma$ is stable and the off diagonal terms represent 
the interaction between the $\sigma$ and the two-particle states, of 
strength $\lambda$. 
To determine the spectrum associated to this matrix we have to diagonalize it;
the eigenvalues of this matrix plotted as functions of $L$ give us the 
spectrum of an interacting toy theory. As an application we used it to plot 
Figure~\ref{spectrum_tot} (Bottom); the ALCs can be seen quite clearly.

\subsection{L\"uscher's method}
\label{luschmethod}

Probably the most well known method of obtaining information on 
resonances on the lattice is the method proposed by 
L\"uscher in Ref~\cite{Luscher:1986pf,Luscher:1991cf}.
The method works by using a mapping which converts information on the 
two-particle spectrum in the elastic region, in a finite volume, 
into information on the scattering phase shift in infinite volume. 
The scattering phase shift will then contain resonance parameters, 
for example near a resonance it will take the form
\begin{eqnarray}
\delta(p) - \frac{\pi}{2} \approx \arctan \left( \frac{4p^{2} + 4M_{\pi}^{2} 
- M_{\sigma}^{2}}  
{M_{\sigma}\Gamma_{\sigma}} \right) \ ,
\label{scattering_total}
\end{eqnarray}
and from this it is possible to extract the resonance mass and width.
Note that, according to Eq.~\ref{scattering_total}, the resonance appears 
when $\delta(p)=0$.

The energies of two-particle states are altered by a finite volume in two ways.
First of all each individual particle in the pair has the virtual polarization
coming from interactions ``around the world'', 
discussed in Ref~\cite{Luscher:1986pf}. 
However there is also a second effect resulting from their direct interaction 
with each other. 
It is this real interaction that L\"uscher's method exploits.

This effect is first derived in the case of non-relativistic quantum mechanics 
and then proven in the field theory case by relating it back to the 
non-relativistic one.
In the quantum mechanical case, the finite volume Schr\"odinger equation, 
provided that the potential has finite range smaller than half the box volume,
has two asymptotic forms.
First, from a scattering theory perspective, since the potential has 
finite range, solutions will have the same asymptotic form as in infinite 
volume and hence will contain the scattering phase shift.
Secondly, outside the potential, the Schr\"odinger equation reduces to the 
Helmholtz equation with eigenvalues given by the energy eigenvalues of the 
two-particle finite volume Hamiltonian. By matching these two asymptotic forms, 
a relationship between the two-particle energies in a finite volume and the 
scattering phase shift in infinite volume is obtained.

In quantum field theory one can decompose the four-point function into an 
infinite sum involving the Bethe-Salpeter Kernel and the two-point function. 
The four point function contains information on the two-particle energy 
spectrum and thus this expansion can be seen as an expansion for the 
two-particle energies. Analytic properties of the Bethe-Salpeter Kernel allow 
the contours of integration in this expansion to be shifted so that the 
two-point propagators take on a non-relativistic form. 
Once expressed this way, the expansion is no different from the Born 
expansion for a non-relativistic theory, with the Bethe-Salpeter Kernel 
filling the r\^ole of a potential. This Born expansion can be seen as 
coming from an effective Schrodinger equation for the two-particle 
wavefunction $\psi(r)$
\begin{equation}
-\frac{1}{2\mu}\nabla\psi(r) + \frac{1}{2}\int{d^{3}r^{'}U_{W}(r,r^{'})}
\psi(r^{'}) = W\psi(r^{'}) \ .
\label{eff_schro}
\end{equation}
The constant $W$ in Eq.~\ref{eff_schro} is the energy, when treated as 
non-relativistic problem, and it is related to the true physical energy in 
the Quantum Field Theory by $E = 2\sqrt{m^{2} + mW}$. The same analytic 
properties mentioned above imply that the Bethe-Salpeter Kernel, as a 
potential, satisfies the conditions on a potential required for the 
quantum mechanical analysis. So the entire framework derived above for 
the non-relativistic case can simply be carried over to quantum field 
theory and L\"uscher's formula holds in this case as well.

This effective Sch\"odinger equation is first constructed in 
Ref~\cite{Luscher:1986pf}.

The relationship derived from this analysis 
is
\begin{align}
\delta(p) &= -\phi(\kappa) + \pi n \ , 
\label{lusherformu}
\\
\tan(\phi(\kappa)) &= \left( \frac{\pi^{3/2}\kappa}{Z_{00}(1;\kappa^{2})} 
\right), \quad \kappa = \frac{pL}{2\pi} \ ,
\label{tanlusherformu}
\end{align}
where $p$ is the relative momentum between the two decay particles. 
$\mathcal{Z}_{js}(1;q^{2})$ is a generalised Zeta function, given by
\begin{equation}
\mathcal{Z}_{js}(1;q^{2}) = \sum_{\underline{n} \in \mathbb{Z}^{3}} 
\frac{r^{j}Y_{js}(\theta,\phi)}{(\underline{n}^{2} - q^{2})^{s}} \ ,
\label{ZETA}
\end{equation}
where $Y_{js}(\theta,\phi)$ are the spherical harmonics. 
Eq.~\ref{lusherformu} is known as L\"uscher's formula. It should be 
mentioned that Eq.~\ref{lusherformu} is in fact a special case of the more 
general expression derived in Ref~\cite{Luscher:1991cf}. The formula 
quoted here is for the spin-$0$ channel, which is the only one relevant 
here. Also in deriving the formula a change in the contour of integration 
allowed the propogators to be rewritten as nonrelativistic propogators. 
However if the volume is quite small this cannot be done because the two-point 
functions will not have the correct initial form due to the polarization 
from around the world as mentioned earlier. For this reason one must 
check that these virtual polarisation effects become negligable 
in order to use L\"uscher's formula.

To obtain resonance parameters using this relationship one proceeds as follows:
\begin{enumerate}
\item Using the Monte Carlo data, obtain the two-particle energy spectrum
$E_n(L)$ as a function of the volume;
\item Using dispersion relations, obtain a momentum from the energy 
spectrum, $p_n(L)$;
\item Compute the appropriate values of $\phi(\kappa)$. Eq.~\ref{lusherformu} 
will then map the values $p_n(L)$ to values of $\delta(p_n(L))$;
\item If this procedure is repated for several energy levels and volumes, 
a profile of $\delta(p)$ is produced;
\item This profile can then be fitted against the Breit-Wigner 
form for $\delta(p)$ in the vicinity of a resonance as given
in Eq.~\ref{scattering_total}.
This fit should give the resonance mass $M_{\sigma}$ and width $\Gamma_{\sigma}$.
\end{enumerate}

\subsection{Histogram method}
\label{subsec_propmeth}
An alternative method of determining the parameters of a resonance is
based on a different way of \emph{analyzing} the finite volume energy
spectrum, Ref~\cite{Bernard:2008ax}. The basic idea is to construct a 
probability distribution $W(E)$ according to the prescriptions:
\begin{enumerate}
\item Measure the two-particle spectrum $E_n(L)$ for different values of $L$
and for $n=1, \cdots, N$ levels;
\item Interpolate the data with fixed $n$ in order to have a continuous
function $E_n(L)$ in an entire range $L \in [L_{0},L_{M}]$;
\item Slice the interval $[L_{0},L_{M}]$ into $M$ equal parts with length
$\Delta{L} = (L_M-L_0)/M$;
\item Determine $E_n(L_i)$ for each $L_i$ ($i=0, \cdots, M$);
\item Choose a suitable energy interval $[E_{min},E_{max}]$ and introduce an
equal-size energy bin with length $\Delta{E}$;
\item Count how many eigenvalues $E_n(L_i)$ are contained in each bin;
\item Normalize this distribution in the interval $[E_{min},E_{max}]$.
\end{enumerate}

The distribution considered in Ref~\cite{Bernard:2008ax} is $W(p)$
but this does not have an important effect on our analysis; 
as a matter of fact, the relation between them is (it is based on the 
definition given in Eq.~\ref{Wp}):
\begin{equation}
W(p)=W(E) \left( \frac{\partial{E}}{\partial{p}} \right) \ ,
\label{wewp}
\end{equation}
where the correct dispersion relation we should use is Eq.~\ref{eqElatt};
the multiplicative term will not modify the Breit-Wigner shape \emph{near}
the resonance.

It is possible to show that the probability distribution $W(p)$ is given
by 
\begin{equation}
W(p)=c \sum_{n=1}^N \left[ p^\prime_n(L) \right]^{-1} 
\label{Wp}
\end{equation}
and differentiating the L\"uscher formula with respect to $L$, it turns out
($c$ is a normalization constant):
\begin{equation}
W(p)=\frac{c}{p} \sum_{n=1}^N \left[L_n(p)+ \frac{2 \pi \delta^\prime(p)}
{\phi^\prime(q_n(p))} \right] \ .
\label{probdistr}
\end{equation}
If one expands $L_n(p)$ around the case of $\delta = 0$, when the there 
is no interaction between the $\sigma$ and the two-particle states, 
this takes the form:
\begin{align}
C^{-1}W(p) = &\sum_{n = 1}^{N} \frac{2\pi}{p^2}\overline{\kappa}_{n} - \\
&\sum_{n = 1}^{N}\frac{2\pi}{p}\frac{1}{\phi^{'}(\overline{\kappa}_{n})}
\left(\frac{\delta(p)}{p} - \delta^{'}(p)\right)  + \mathcal{O}(\delta^{2}) \ .
\label{histdef2}
\end{align}
The first term is equivalent to the histogram that would be constructed 
in a theory where the $\sigma$ is a stable particle.
We will call this histogram the free background, $C_{0}^{-1} W_{0}(p)$, 
where $C_{0}$ is its normalisation constant. If we subtract it from 
the interacting histogram we obtain
\begin{align}
C^{-1}W(p) - C_{0}^{-1} W_{0}(p) =& &\\ 
- \sum_{n = 1}^{N} \frac{2\pi}{p}\frac{1}{\phi^{'}(\overline{\kappa}_{n})}
\left(\frac{\delta(p)}{p} - \delta^{'}(p)\right) + \mathcal{O}(\delta^{2}) \ .
& &
\label{histdef3}
\end{align}

In the limit of an infinite number of energy levels, i.e. infinite volume, 
the terms of $\mathcal{O}(\delta^{2})$ are very small for the
vast majority of energy levels and so become negligable. Hence we obtain:
\begin{align}
C^{-1}W(p) - C_{0}^{-1} W_{0}(p) \approx & \\
- \left[\sum_{n = 1}^{N} \frac{2\pi}{\phi^{'}(\overline{\kappa}_{n})}\right]
\frac{1}{p}\left(\frac{\delta(p)}{p}  - \delta^{'}(p)\right) \ . &
\label{histdef4}
\end{align}
However we can see that
\begin{equation}
\sum_{n = 1}^{N} \frac{2\pi}{\phi^{'}(\overline{\kappa}_{n})}
\end{equation}
is just a constant independent of $\delta$ or $p$ and so it can just be 
absorbed into the normalisation constant of the histogram to give us:
\begin{equation}
W(p)-W_0(p) \propto \frac{1}{p} \left( \frac{\delta(p)}{p} -
  \delta'(p) \right) \ .
\label{wpmwp0}
\end{equation}

This last quantity is determined by $\delta(p)$ alone and close to the
resonance, assuming a smooth dependence on
$p$ for the other quantities, it follows the Breit-Wigner shape of the
scattering cross section with the same parameters:

\begin{equation}
W(p)-W_0(p) \propto \frac{1}{[E(p)^2-M_\sigma^2]^2+M_\sigma^2 \ \Gamma^2} \ .
\label{bw}
\end{equation}
To emphasize the approximations that are present we note that, 
because
\begin{equation}
\delta(p)=\arctan\left(\frac{M_\sigma \Gamma}{M_\sigma^2-E^2(p)}\right) \ ,
\label{delt1}
\end{equation}
then we can work out:
\begin{equation}
\delta^\prime(p)=\frac{8 M_\sigma \Gamma p}{[M_\sigma^2-E^2(p)]^2+M_\sigma^2
\Gamma^2 } \ ;
\label{delt2}
\end{equation}
therefore we can see that the Breit-Wigner shape of Eq.~\ref{bw} is entirely
due to $\delta^\prime(p)$. It should also be noted that this Histogram 
method does no require one to have knowledge of the function $\phi(\kappa)$, 
unlike like L\"uscher's method.

In Ref~\cite{Bernard:2008ax} this method is tested on \emph{synthetic} 
data produced using the L\"uscher formula by experimentally measured phase 
shifts and in Ref~\cite{Morningstar:2008mc} it is tested on nonrelativistic 
quantum mechanics.
The main task of our work is to test this method on an effective field theory
where a resonance emerges, producing data by lattice simulations.

\section{The $O(4)$ sigma model}
\label{themodel}
The model we have used in our simulations is the $O(4)$ model
in the broken phase. This model has previously been used to test L\"uscher's 
method, Ref~\cite{Gockeler:1994rx}. The Lagrangian is the following
(with $i=1,2,3,4$):
\begin{equation}
\mathcal{L}=\onehalf \partial \phi_i \partial \phi_i + \lambda
( \phi_i^2-\nu^2 )^2-m^2_{\pi,0} \nu \phi_4 \ .
\label{lagr0}
\end{equation}
The term proportional to $\phi_4$ is introduced to break the
symmetry explicitly in order to give mass to the three Goldstone bosons.
To understand the meaning of the terms and the parameters in the Lagrangian,
we first introduce the new fields $\sigma$ and $\rho_i$
(with the constraint $\rho_i \rho_i=1$):
\begin{equation}
\phi_i=(\nu+\sigma) \rho_i \ ,  \qquad \mbox{with $i=1,2,3,4$} \ ;
\label{newfileds}
\end{equation}
then, we expand the potential around
the classical minimum $\phi_i\phi_i =\nu^2$
(using also $\rho_i \partial \rho_i=0$):
\begin{eqnarray}
\mathcal{L} = \onehalf \nu^2 \partial \rho_i \partial \rho_i +
\onehalf \sigma^2 \partial \rho_i \partial \rho_i +
\nonumber \\
\onehalf \partial \sigma \partial \sigma +
\nu \sigma \partial \rho_i \partial \rho_i +
\lambda \sigma^4 +4\nu^2 \lambda \sigma^2+
\nonumber \\
4 \nu \lambda \sigma^3
-m_{\pi,0}^2 \nu^2 \rho_4-m_{\pi,0}^2 \nu \sigma \rho_4 \ .
\label{lagr1}
\end{eqnarray}
The $\sigma$ field is clearly related to the massive $\phi_4$ field in the
original Lagrangian, whereas the four constrained fields $\rho_i$ are related to the 
three ``pions''. In the form Eq.~\ref{lagr1} we can not directly interpret the physical
content of the Lagrangian, due to the presence of linear terms. Particularly 
it is not obvious that the explicit breaking term has given the three Goldstone bosons a mass.

There is an easy way to rewrite the Langrangian to make all of this more obvious, based 
on the treatment of the non-linear sigma model (see for example Ref~\cite{DeGrand:2006zz}, Sec~2.4.3.).

We introduce the pions using an element of $SU(2)$:
\begin{equation}
U=e^{\frac{i}{\nu}\pi_j \sigma_j }
= \cos{\left( 
\frac{|\vec\pi|}{\nu} \right)} + i \sigma_j \frac{\pi_j}{|\vec{\pi}|} 
\sin{\left( \frac{|\vec\pi|}{\nu} \right)}   \ , 
\label{Uexpr1}
\end{equation}
where $\sigma_j$ are the three Pauli matrices and where $\nu$ plays the r\^ole of the pion decay constant. 
On the other hand, we can also form $SU(2)$-valued fields from our 
$\rho$-fields by
\begin{equation}
U=\rho_4+i \sigma_j \rho_j \ ,
\label{Uexpr2}
\end{equation}
with $j=1,2,3$ and the constraint $\rho_4^2+\rho_j \rho_j =1$.
We can therefore identify the connection between the three fields $\pi_j$ and 
the fourth $\rho_j$, using Eq.~\ref{Uexpr1} and Eq.~\ref{Uexpr2}:
\begin{eqnarray}
\rho_4 &=& \cos{\left( \frac{|\vec\pi|}{\nu} \right)} \ , \label{rho4cos}\\
{\rho}_j &=& \frac{\pi_j}{|\vec{\pi}|} 
\sin{\left( \frac{|\vec\pi|}{\nu} \right)} \ .
\end{eqnarray}
We can now replace the $\rho$ fields in the Langrangian using 
%
the expression $\frac{1}{2}\Tr(\partial_{\mu}U\partial_{\mu}U^{\dagger})$. 
For the $\rho$ fields this is
\begin{align}
\frac{1}{2}\Tr(\partial_{\mu}U\partial_{\mu}U^{\dagger}) = \sum_{i = 1}^{4} \partial_{\mu}\rho_{i}\partial_{\mu}\rho_{i} \ .
\end{align}
For the pion fields this gives
\begin{align}
\frac{1}{2}\Tr(\partial_{\mu}U\partial_{\mu}U^{\dagger}) = \frac{1}{\nu^2}\sum_{i = 1}^{3} \partial_{\mu}\pi_{i}\partial_{\mu}\pi_{i} \ ,
\end{align}
and so we have:
\begin{align}
\sum_{i = 1}^{4} \partial_{\mu}\rho_{i}\partial_{\mu}\rho_{i} = \frac{1}{\nu^2}\sum_{i = 1}^{3} \partial_{\mu}\pi_{i}\partial_{\mu}\pi_{i} \ .
\label{rhoder}
\end{align}
Eq.~\ref{rho4cos} and Eq.~\ref{rhoder} can then be substituted into the 
original Lagrangian, Eq.~\ref{lagr1}. Expanding the $\cos\left(\frac{|\vec\pi|}{\nu}\right)$, which has replaced the $\rho_{4}$ field, we obtain as our Lagrangian:
\begin{eqnarray}
\mathcal{L}=\frac{1}{2} \partial_{\mu} \pi_j \partial_{\mu} \pi_j +
\frac{1}{2\nu^2} \sigma^2 \partial_{\mu} \pi_j \partial_{\mu} \pi_j +
\onehalf \partial_{\mu} \sigma \partial_{\mu} \sigma +
\nonumber \\
\frac{1}{\nu} \sigma \partial_{\mu} \pi_j \partial_{\mu} \pi_j +
\lambda \sigma^4 +4\nu^2 \lambda \sigma^2+4 \nu \lambda \sigma^3+
\nonumber \\
\frac{1}{2} m_{\pi, 0}^2 \pi_j \pi_j+
\frac{m_{\pi, 0}^2}{2\nu} \sigma \pi_j \pi_j \ + \ldots \ , \quad
\end{eqnarray}
where the higher order terms include higher order couplings between the pions and the $\sigma$ resonance, as well as pion self-interaction terms.
We can see that the $\sigma$ field gets a bare mass 
\begin{align}
m_\sigma=2 \nu \sqrt{2 \lambda} \ ,
\label{lagrmass}
\end{align}
and due to terms such as the three-point interaction $\nu \sigma \partial \rho_i \partial \rho_i$ the sigma particle is unstable. We can also see that the parameter $m_{\pi, 0}$, that we introduced in Eq.~\ref{lagr0}, functions as the bare pion mass. So our 
explicit soft-breaking term has given the Goldstone bosons a mass.

Two things should be noted about the three-point interaction term. 
First of all, it depends on $\nu$, so the
sigma resonance should be broader with decreasing values of $\nu$. 
We will not however make direct use of this, since making $\nu$ too small leads
to symmetry restoration.
The interaction also contains a derivative. In momentum space this will give 
an extra $p^2$ term to the vertex appearing in Feynman diagrams. We expect 
the interaction between the pions and the sigma resonance to be stronger when 
the pions have larger relative momentum.
The decay rate of the sigma resonance will also depend on $\lambda$, since the $\sigma$ field self-coupling terms will affect 
the interactions between the $\sigma$-particle and the pions.

For certain values of the parameters the $O(4)$ symmetry will be restored and 
the theory will enter the unbroken phase.  Since we do not want this to occur 
we must avoid the region of the $\lambda$, $\nu$ parameter space in which the 
symmetry remains unbroken. For any fixed value of $\lambda$ the symmetry is 
restored when $\nu$ is sufficiently small. The point of this phase 
transition $\nu_{*}(\lambda)$ increases with increasing $\lambda$. In 
particular
\begin{equation}
\lim_{\lambda \rightarrow \infty} \nu_{*}(\lambda) \approx 0.78 \ .
\label{brokenlim}
\end{equation}
Hence we will always keep $\nu$ above $0.78$, specifically we use $\nu = 1$ or $1.05$, to guarantee that the symmetry remains
broken.

A derivation of Eq.~\ref{brokenlim} is contained in Ref~\cite{Gockeler:1994rx}, although there, due to different parameters being
used, it appears as $\kappa_c \approx 0.304$.

\section{Monte Carlo simulation}
\label{secmontecarlo}
The theory described by the Lagrangian in Eq.~\ref{lagr0}, was simulated using
an over-relaxation algorithm for the first three near-Goldstone fields, 
followed by a Metropolis update to guarantee the ergodicity, and a Metropolis
algorithm for the massive field, $\phi_4$.

In order to determine the single particle spectrum we first introduce the
partial Fourier transform (PFT) of the four fields $\phi_i$:
\begin{equation}
\tilde{\phi_i}(\vec{n},t)=\frac{1}{V}\sum_x \phi_i(\vec{x},t) 
e^{-i \vec{x} \vec{p}} \ , \qquad
p_i=\frac{2 \pi}{L_i} n_i \ , 
\end{equation}
where $n_i=0, \ldots ,L_i-1$.
The single particle mass is extracted from the zero momentum correlation
function ($\vec{n}=\vec{0}$):
\begin{equation}
C_i(t)= \langle \tilde{\phi_i}(\vec{n},t) \tilde{\phi_i}(-\vec{n},0) \rangle \ .
\end{equation}
In particular with $i=1,2,3$ we can determine the mass of the three pion fields;
with $i=4$ we extract the mass of the $\sigma$ particle.
In Monte Carlo simulation, the mass of the lightest state is usually 
determined with better statistical precision, and this is observed here, where
$m_\pi$ is determined with a higher precision then $m_\sigma$;
this is not a problem because we are mainly interested
in a good resolution of states consisting of two pions and these energies are
well determined. 

The two-particle spectrum is measured by introducing operators with
zero total momentum and zero isospin:
\begin{equation}
O_{\vec{n}}(t)= \sum_{i=1}^3 \tilde{\phi_i}(\vec{n},t)
\tilde{\phi_i}(-\vec{n},t) \ ;
\end{equation}
we take into account $N-1$ different operators corresponding to
$n^2=0,1,\ldots,N-2$.
An $N$-th operator, that clearly has the correct quantum number is the PFT
of the field $\sigma$ (actually of $\phi_4$) with $\vec{p}=0$.
\begin{figure}[htb]
\begin{center}
\hspace{-2mm}
   \includegraphics[width=0.36\textwidth,angle=-90]{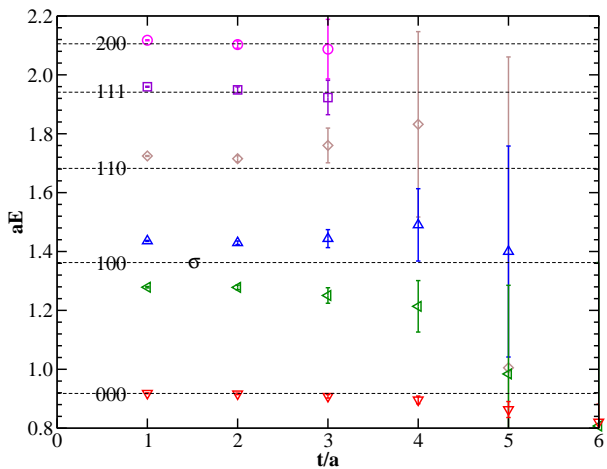}
\caption{Effective mass versus time as determined by the diagonalization
of the correlation matrix $C_{ij}$. The dashed constant lines describe the 
free two-particle spectrum. Simulation parameters: 
$\nu=1.0$, $\lambda=1.4$, $am_\pi=0.36$, volume=$12^3 \times 64$.}
\label{L12masseff}
\end{center}
\end{figure}
To determine the energy levels we use a method, introduced
in Ref~\cite{Michael:1985ne} (see also Ref~\cite{Luscher:1990ck}), based on a generalized eigenvalue problem applied
to the correlation matrix function $C_{ij}(t)= \langle O_i O_j\rangle$,
that is a matrix whose elements are all possible correlators between the $N$
operators:
\begin{equation}
C(t) \psi = \lambda(t,t_*) C(t_*) \psi \ ,
\end{equation}
where $t_*$ is fixed to a small value (we verified that in this model
the results are insensitive to its value, so we chose $t_*=0$). It is possible
to show that the eigenvalues, for $\alpha=1, \ldots, N$, behave as
\begin{equation}
\lambda_\alpha(t,t_*)=e^{-(t-t_*)E_\alpha} \ ,
\end{equation}
where $E_\alpha$ describes the spectrum of the theory;
a typical result is shown in Figure~\ref{L12masseff}. Using this method
we can see a wide plateau of approximately six lattice spacings for the 
ground-state, dominated by two pions at rest that starts from $t_0=1$ in this 
case.  The width of the plateau decreases with increasing energy and 
it is just 2 lattice spacing for the level $(2,0,0)$ the onset of the plateau 
also occurs later.
In Fig.~\ref{L12masseff}, it is evident there is strong 
mixing between the state resembling two pions, each with momentum 
$n=\pm(1,0,0)$ and the $\sigma$ state, illustrating an example avoided level 
crossing. 

\subsection{Histogram Results}
\label{subsechisres}

\begin{figure}[htb]
\hspace{-2mm}
  \includegraphics[width=0.36\textwidth,angle=-90]{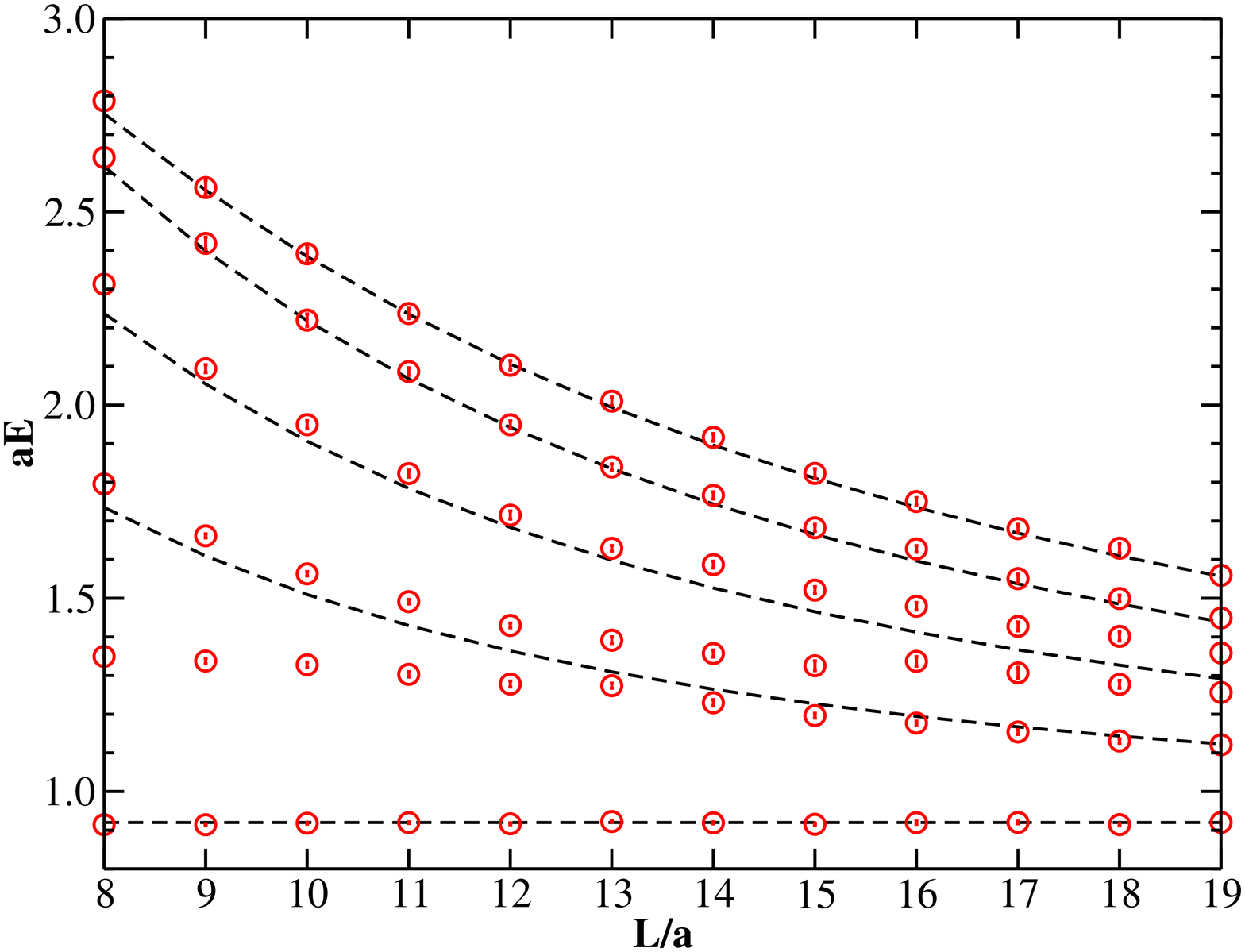}
\hspace{2mm}
  \includegraphics[width=0.36\textwidth,angle=-90]{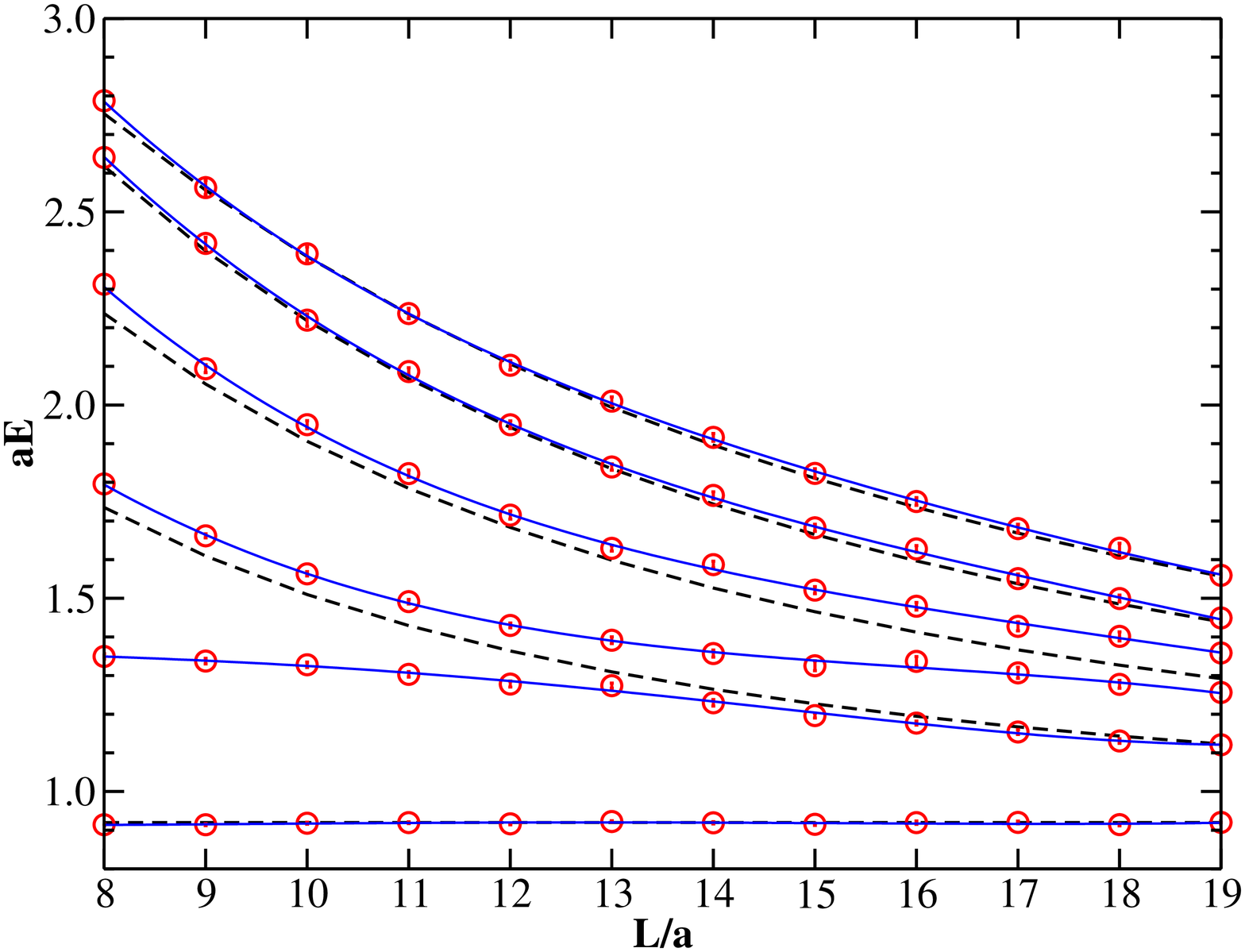}
\caption{(Top) Spectrum of the theory for different values of the
volume for the following simulation parameters:
$\nu=1.0$, $\lambda=1.4$, $am_{\pi,0}=0.36$.
The dashed lines describe the free two-particle spectrum. 
(Bottom) The interpolated data using a polynomial.}
\label{L12spectr}
\end{figure}

In order to test the applicability of the two methods 
for different widths of resonance, we consider three different sets of 
parameters in the lagrangian.
In all three simulations, the time extent of the lattice is fixed to $L=64a$. 
The first simulation is performed using $\nu=1.0$, $\lambda=1.4$, 
$am_{\pi,0}=0.36$.
These parameters were determined to have the intersection between the $\sigma$ 
energy level and $n=(1,0,0)$ two-particle energy level in the absence of 
interaction at around $L=12a$. The measured mass for the pion turns out to be 
$am_\pi=0.460(2)$. The first six energy levels were determined clearly for 
different volumes in the range $8 \leq L/a \leq 19$. The fractional error 
on the measured energies was in the range 0.5\% - 1.0\%. 
The top panel of Figure~\ref{L12spectr} shows the results of this set of  
simulations.  Each energy was determined from statistical fitting, choosing 
the onset of the plateau to be $t_0=2$. 
Constructing a histogram where a resonance is clearly seen requires a large
set of lattice volumes and energy levels. We found the stability of the
histogram could be enhanced by interpolating the spectrum data using 
polynomials in $L$ over all values of $L/a$ in the measured range and using 
these polynomials to generate more data for the histogram. 
The lower panel of Fig.~\ref{L12spectr} shows the resulting polynomial fit. 
Polynomials of order 3, 4 and 5 were used, which provided a means of evaluating
the systematic errors in the final results. 
The dashed lines in Fig.~\ref{L12spectr} show the free two-particle spectrum,
calculated using Eq.~\ref{eqElatt}.  

In order to control the dominant distortions in the free spectra arising 
simply from discretisation artefacts for these high-lying states, 
the energy curves for the non-interacting pions were computed using the free
dispersion relation after first computing the subtracted mass $m_{\pi,r}$ 
using the rest-energy of a single pion $m_\pi$:
\begin{equation}
m_\pi=4 \sinh^{-1}\left[{\onehalf m_{\pi,r} }\right] \ .
\end{equation}
These curves show very good agreement
with the observed spectra away from the resonance even at very high energies.
The pion mass measured in these simulations is $am_\pi=0.460(2)$, which 
differs from the parameter in the lattice lagrangian $am_{\pi,0}=0.36$ due
to renormalisation effects of the interacting theory. 
%
\begin{figure}[htb]
\begin{center}
  \includegraphics[width=0.36\textwidth,angle=-90]{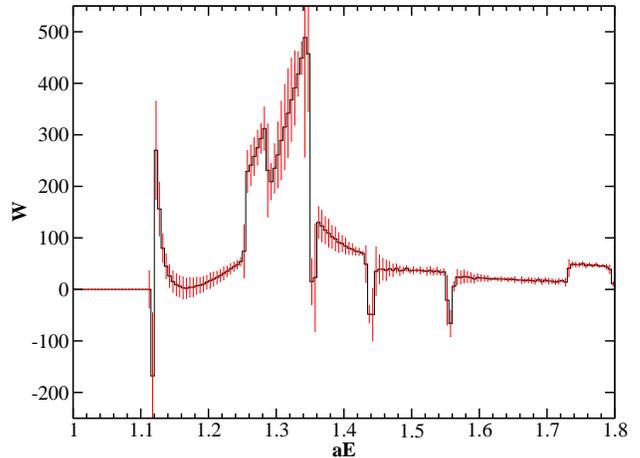}
\caption{The probability distribution $\tilde{W}=W-W_0$ obtained from data from
Figure~\protect\ref{L12spectr}.}
\label{L12histdiff}
\end{center}
\end{figure}
%
This free spectrum is used to determine the distribution $W_0(p)$ which is 
then subtracted from $W$, obtained from the interacting spectrum.
It is important to note that if $N$ levels are used to plot $W$ in the 
interacting spectrum, then the number of levels needed from the free spectrum 
to determine $W_0$ is $N-1$. This takes into account the extra level arising 
from the resonance and ensures the correct modes are subtracted at the upper 
and lower ends of the energy range. 
Note that, in the general case, if the same number $N$ of levels are used 
the final result will be affected by the presence of unwanted peaks which
potentially could hidden the presence of the resonance.

Using polynomials, we are able to produce a large number of data at arbitrary
values of $L/a$; we use a resolution of $a\Delta{L}=0.001$.
Using a bin width of $a\Delta{E}=0.005$, we get the probability distribution 
$W$, described in Sec~\ref{subsec_propmeth}, and, consequently, the histogram 
$\tilde{W}=W-W_0$ of Figure~\ref{L12histdiff}. 
Note that to get $\tilde{W}$ both $W$ and $W_0$
are determined from the same range of $L/a \in [8,19]$.
The error bars in Figure~\ref{L12histdiff} then include both the systematic
(determined by the different results we get using the different polynomials) 
and statistical errors coming from the histogram $W$ and the statistical 
errors coming from the determination of $W_0$ including the statistical error 
propagating from $m_\pi$ via Eq.~\ref{eqElatt}.

Clearly, the shape of the histogram in Figure~\ref{L12histdiff} is far from 
a Breit-Wigner distribution. The dominant reason for the distortions is that
our Monte Carlo measurement determined only six energy levels while the 
conclusions of Sec~\ref{subsec_propmeth} are true only in the limit of an 
infinite number of levels. Many jumps and spikes are seen. Our task is now to 
try to modify the analysis in order to get fewer artefacts from the same raw 
data.
We investigated the origin of the spikes and concluded that they are
related to subtractions of an ``incorrect'' background $W_0$. It is easy to see
that the spikes appear every time there is the intersection between the
six levels of the interacting spectrum or of the five levels of the free
spectrum with the extremities of the volume range at $L/a=8$ and $L/a=19$.
Near those two extremities a more careful modelling of the free background is
needed;
\begin{figure}[htb]
\hspace{-2mm}
  \includegraphics[width=0.36\textwidth,angle=-90]{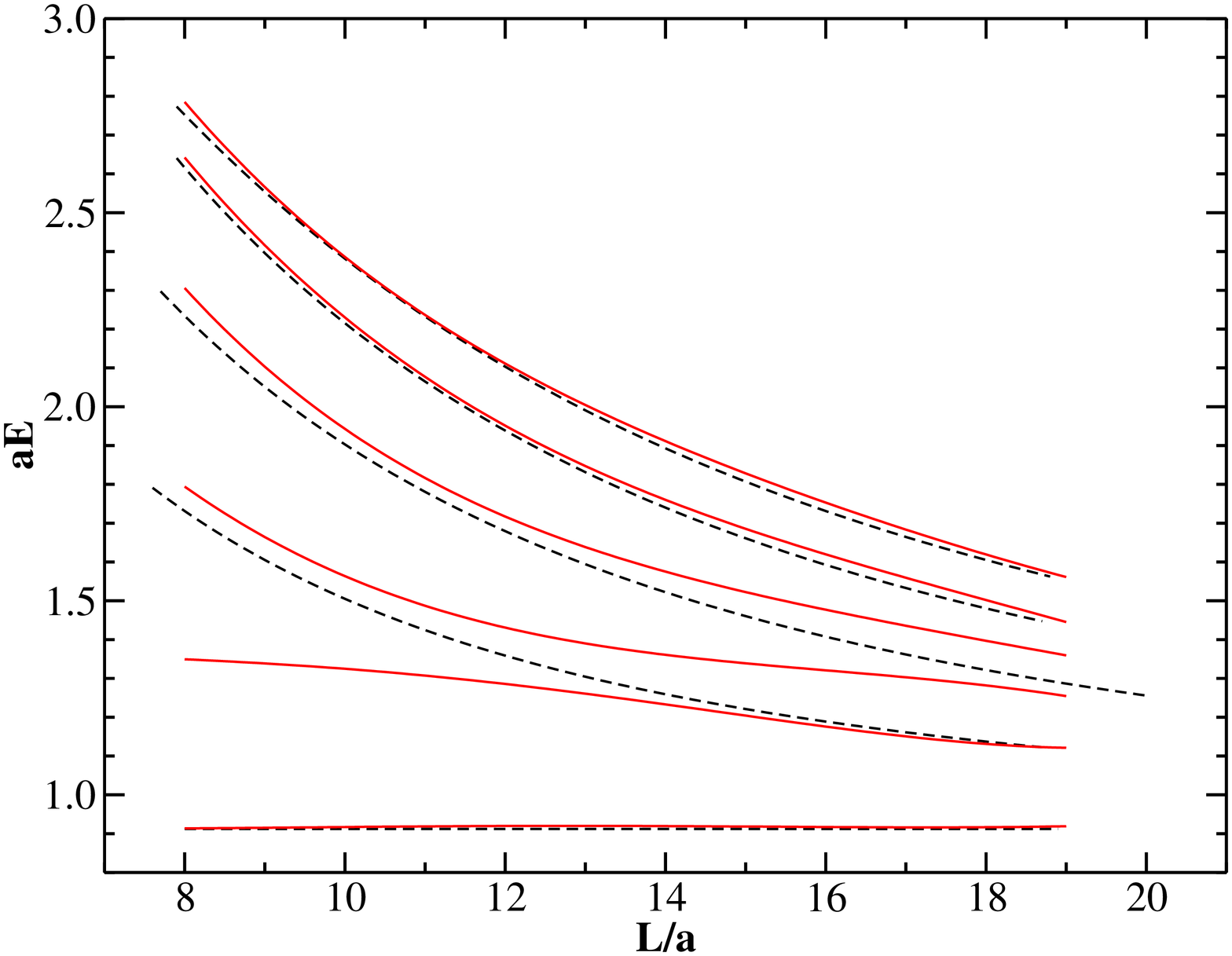}
\hspace{2mm}
  \includegraphics[width=0.36\textwidth,angle=-90]{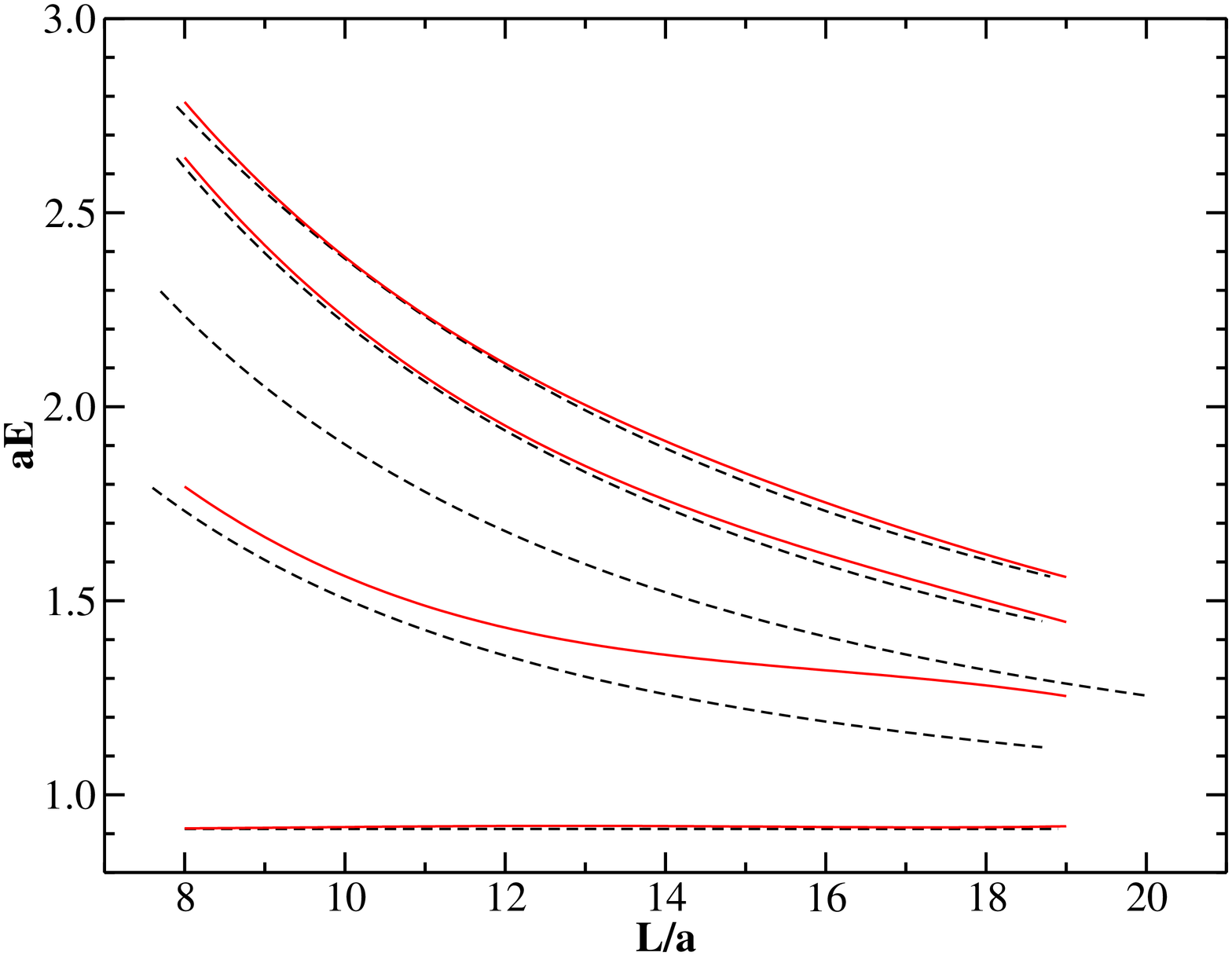}
\caption{(Top) Energy levels of Figure~\protect\ref{L12spectr} (Bottom) 
with the correct free two-particle spectrum background. 
(Bottom) Energy levels where we deleted the two levels that appear without their 
own background.}
\label{L12spectrcorr}
\end{figure}
Figure~\ref{L12spectrcorr}~(Top) shows a corrected background subtraction. 
In order to correctly subtract the free background, each free spectrum line is
extended using the polynomial fitting form. This is done so that the extremity 
of each line has energy equal to the value at the end of the interacting 
spectrum line closest to it. In this way all interacting lines are subtracted
correctly rather than the subtraction being affected by the limit
of the volume range that we are actually using in our simulations.
Using this procedure to determine $W_0$ we get the \emph{correct}
histogram of Figure~\ref{L12histmod} (Top).
\begin{figure}[htb]
\hspace{-2mm}
  \includegraphics[width=0.36\textwidth,angle=-90]{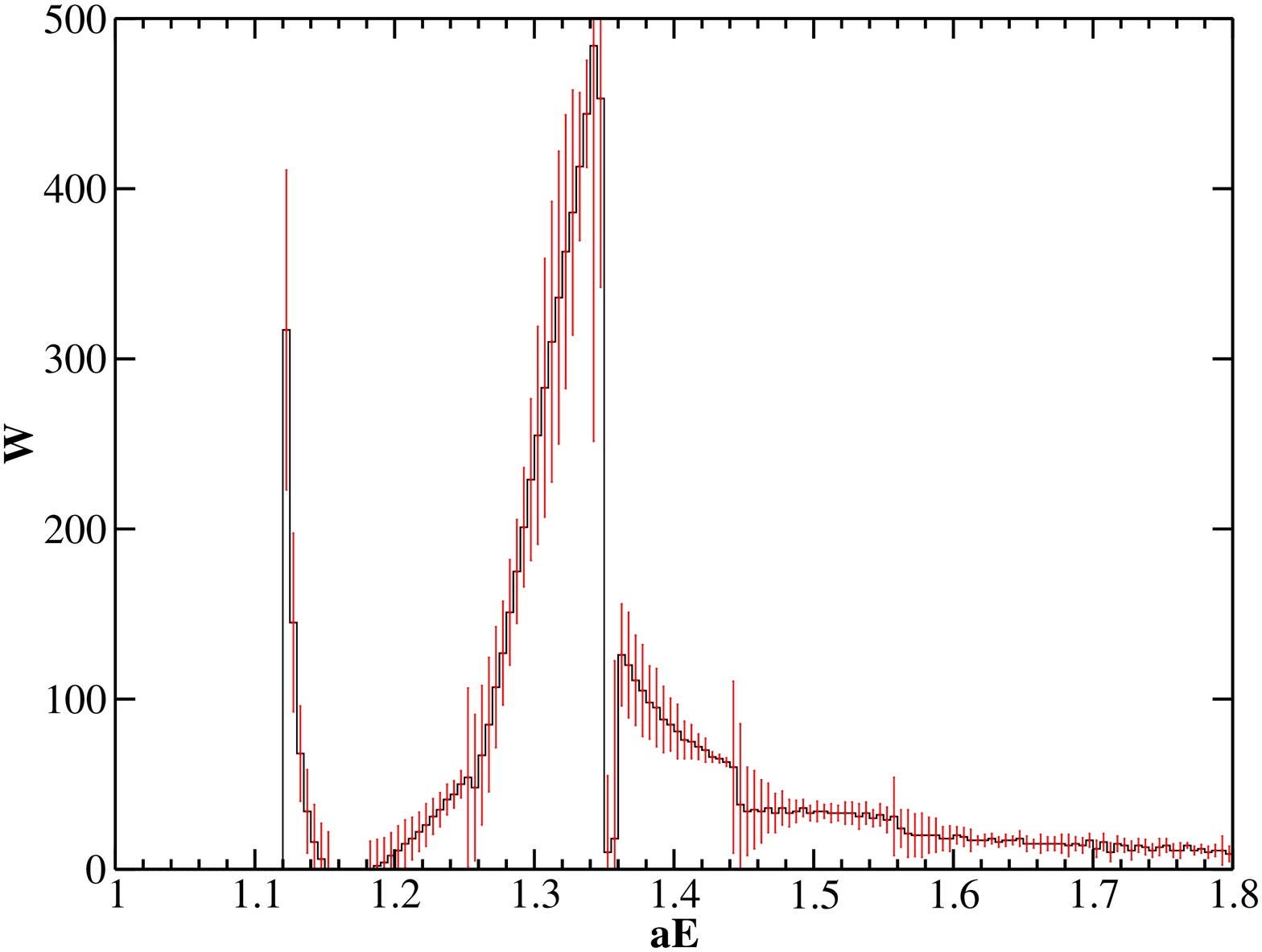}
\hspace{2mm}
  \includegraphics[width=0.36\textwidth,angle=-90]{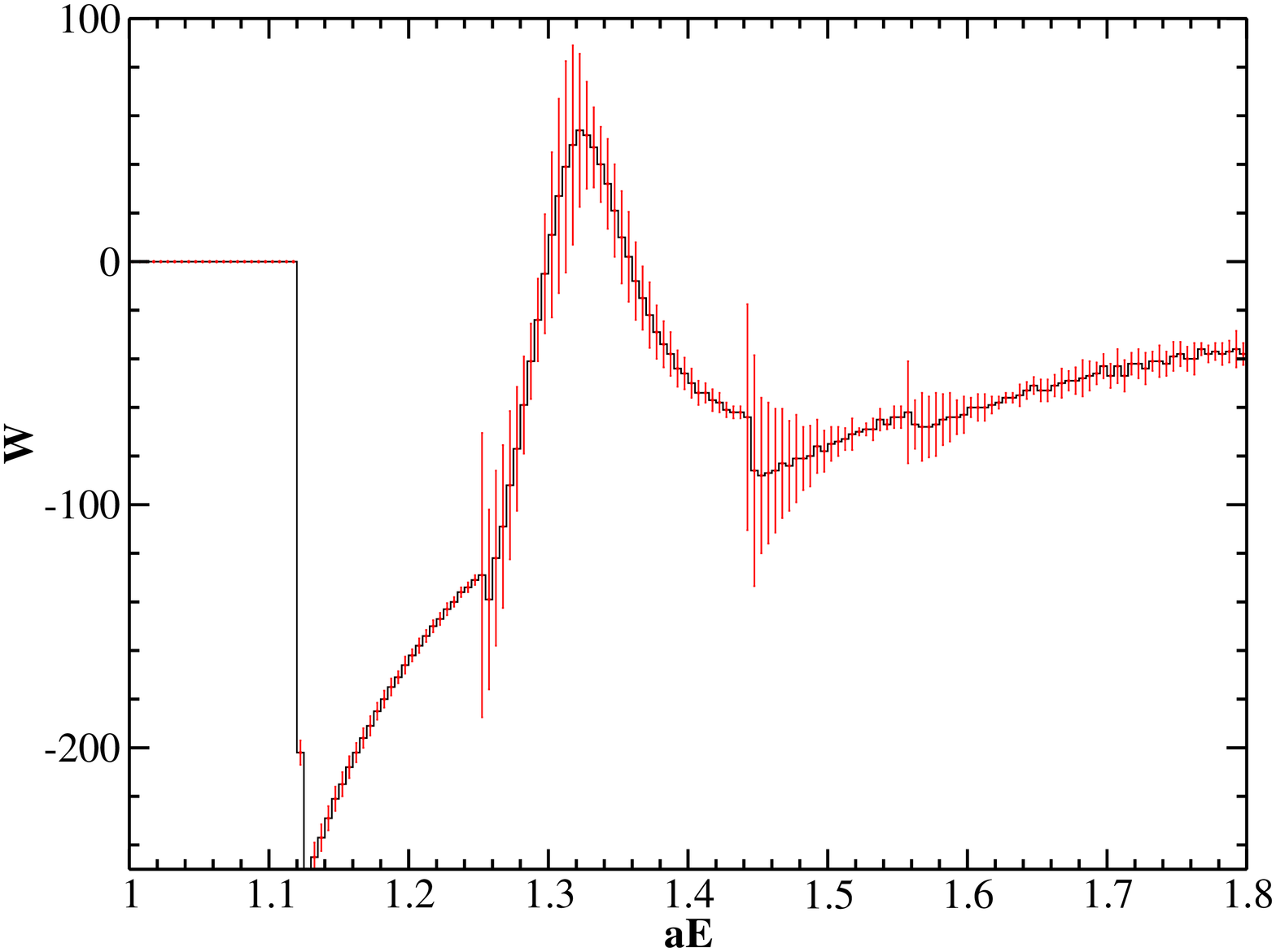}
\caption{(Top) Probability distribution $\tilde{W}$ obtained by data from
Figure~\protect\ref{L12spectrcorr} (Top).  (Bottom) Probability distribution 
$\tilde{W}$ obtained from Figure~\protect\ref{L12spectrcorr} (Bottom).}
\label{L12histmod}
\end{figure}
Unfortunately, in Figure~\ref{L12histmod} (Top), we continue to see a 
discontinuity at $aE \approx 1.35$; the origin of this can be understood by 
looking at Figure~\ref{L12spectrcorr} (Top). There are two extremity lines, 
one at $L=8a$ and one at $L=19a$ which are both around $E \approx 1.35$, that 
occur without a corresponding ``background''; actually, in this case the 
background is the resonance itself we are looking for.

Therefore, there is no way to avoid the presence of this jump because
we do not know anything about the resonance; the only thing we can do is
to completely exclude from our analysis those two levels, 
Figure~\ref{L12spectrcorr} (Bottom), hoping that
the resonance still appears in other modes. In Figure~\ref{L12histmod} 
(Bottom) we show the probability distribution $\tilde{W}$ in this last case; 
now clearly a Breit-Wigner distribution emerges.

\begin{figure}[htb]
\begin{center}
\hspace{-2mm}
  \includegraphics[width=0.36\textwidth,angle=-90]{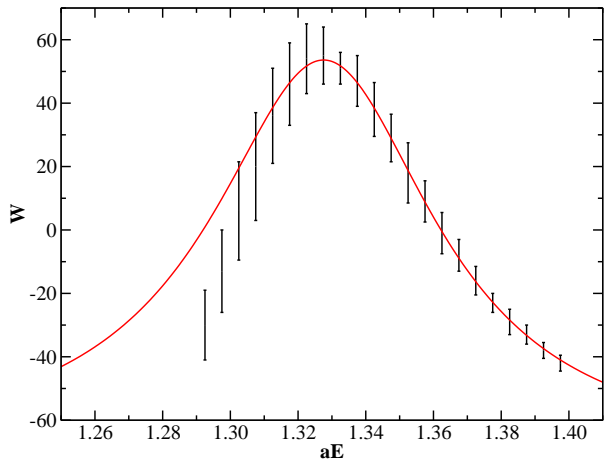}
\caption{Data from Figure~\ref{L12histmod} (Bottom) that we fitted to 
determine the resonance parameters with the final curve fitting.}
\label{L12fit}
\end{center}
\end{figure}

\begin{figure}[htb]
\hspace{-2mm}
  \includegraphics[width=0.36\textwidth,angle=-90]{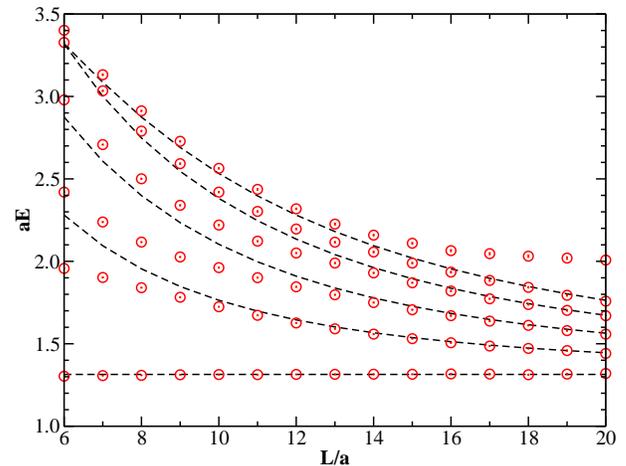}
\hspace{2mm}
  \includegraphics[width=0.36\textwidth,angle=-90]{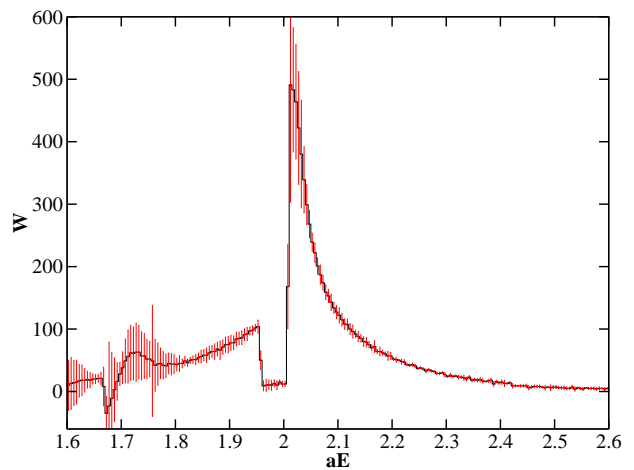}
\caption{(Top) Spectrum of the theory for different values of the
volume for the following simulation parameters: $\nu=1.0$, $\lambda=4.0$,
$am_{\pi,0}=0.56$.  (Bottom) The probability distribution considering 
the correct background.}
\label{L8a}
\end{figure}

It is now possible to fit these data to Eq.~\ref{bw} to determine the
parameters of the resonance, Figure~\ref{L12fit}. Applying a sliding 
window procedure around the peak gives: 
$aM_\sigma=1.330(5)$ and $a\Gamma_\sigma=0.10(5)$.
We simulated the theory at a second set of parameters corresponding
to a broader resonance:
$\nu=1.0$, $\lambda=4.0$, $am_{\pi,0}=0.56$. In this case, the parameters were 
chosen such that the intersection occurs between the $\sigma$ energy level and
$n=(1,0,0)$ two-particle energy level close to $L=8a$. The measured mass
for the pion turns out to be $am_\pi=0.657(3)$.
In Figure~\ref{L8a} (Top) we plot the spectrum for $6 \leq L/a \leq 20$
for the first six levels; in this case the onset value for the plateaux 
is $t_0=1$ and the relative error varies in the range 0.05\% - 0.2\%.

\begin{figure}[htb]
\hspace{-2mm}
  \includegraphics[width=0.36\textwidth,angle=-90]{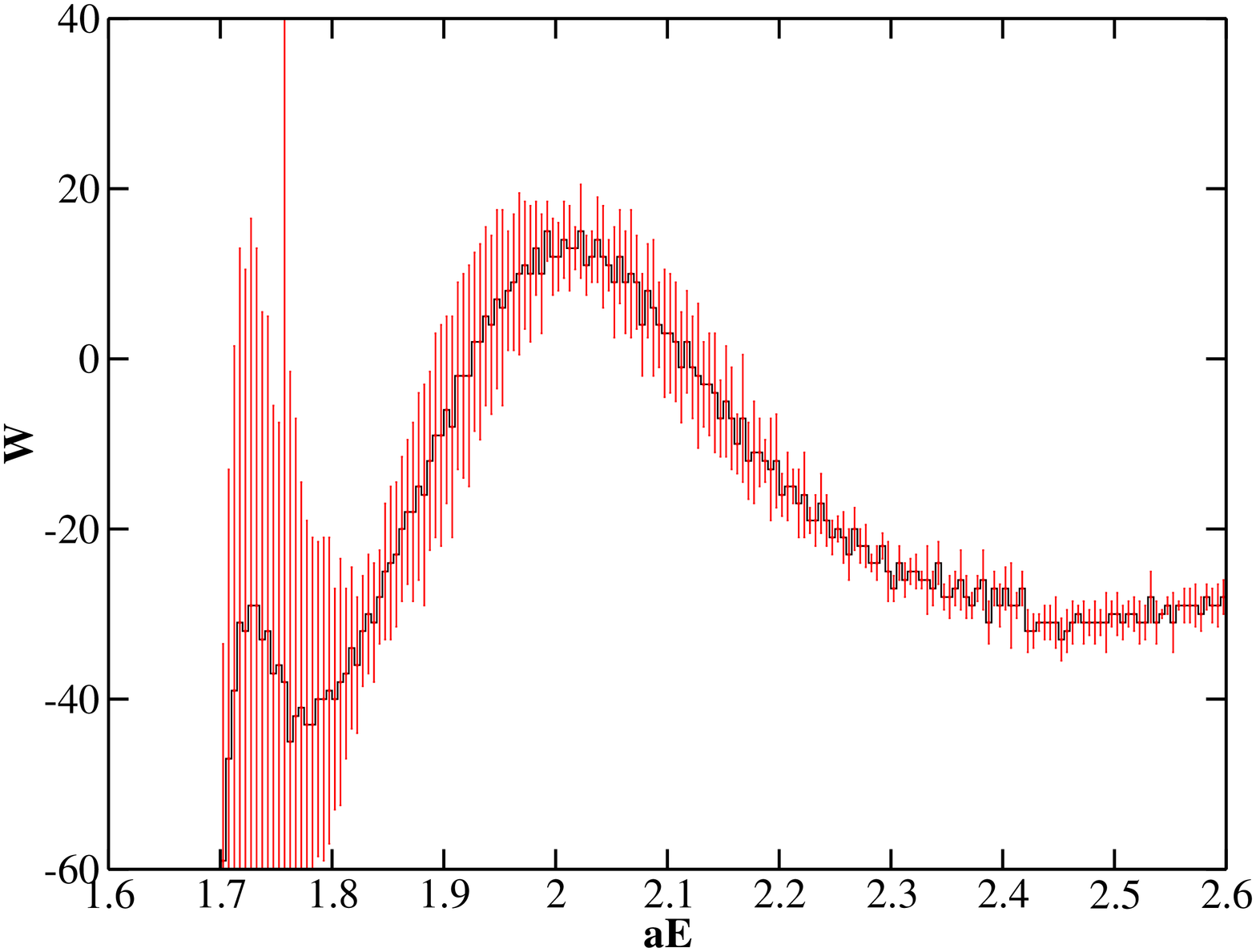}
\hspace{2mm}
  \includegraphics[width=0.36\textwidth,angle=-90]{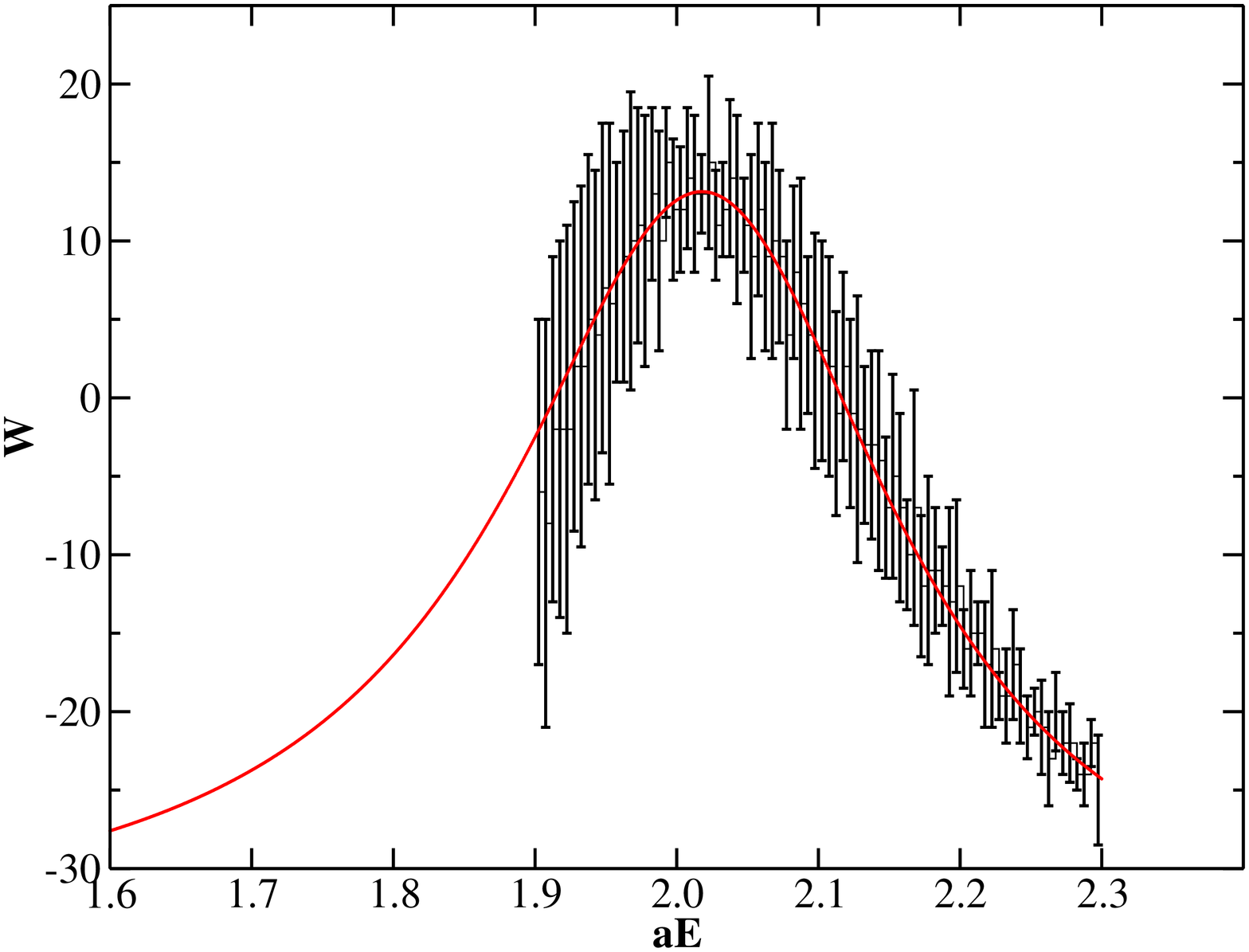}
\caption{(Top) Probability distribution $\tilde{W}$ using the correct
background and excluding the two levels that are without a corresponding
background.  (Bottom) Data we fitted to determine the resonance 
parameters with the final curve fitting. Simulation parameters: 
$\nu=1.0$, $\lambda=4.0$, $am_{\pi,0}=0.56$.}
\label{L8c}
\end{figure}

\begin{figure}[htb]
\begin{center}
\hspace{-2mm}
  \includegraphics[width=0.36\textwidth,angle=-90]{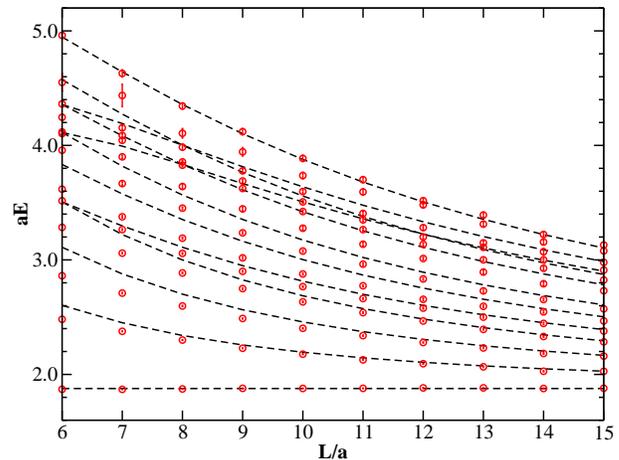}
\caption{Spectrum of the theory with simulation parameters:
 $\nu=1.0$, $\lambda=200.0$, $am_{\pi,0}=0.86$.}
\label{L10a}
\end{center}
\end{figure}

\begin{figure}[htb]
\hspace{-2mm}
  \includegraphics[width=0.36\textwidth,angle=-90]{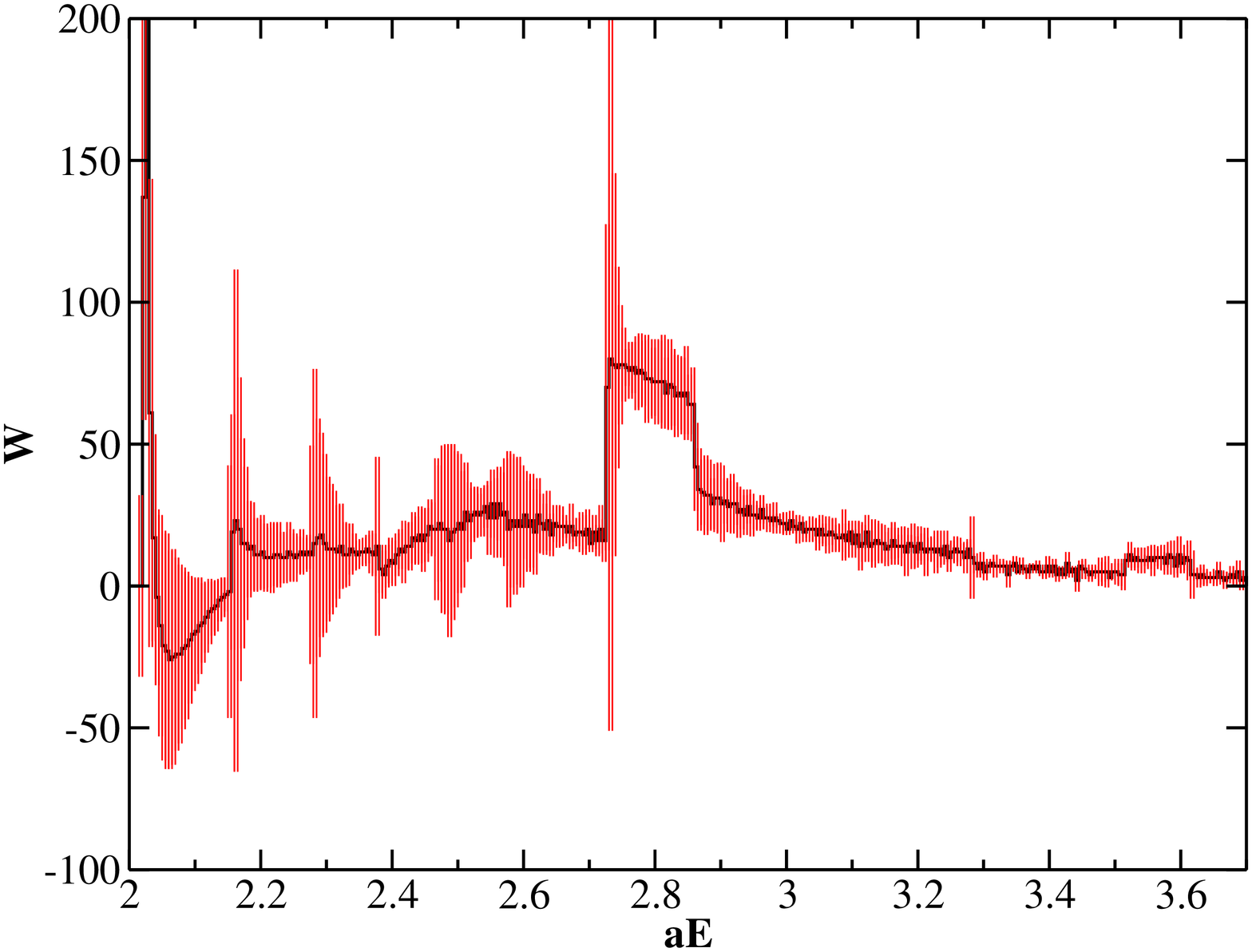}
\hspace{2mm}
  \includegraphics[width=0.36\textwidth,angle=-90]{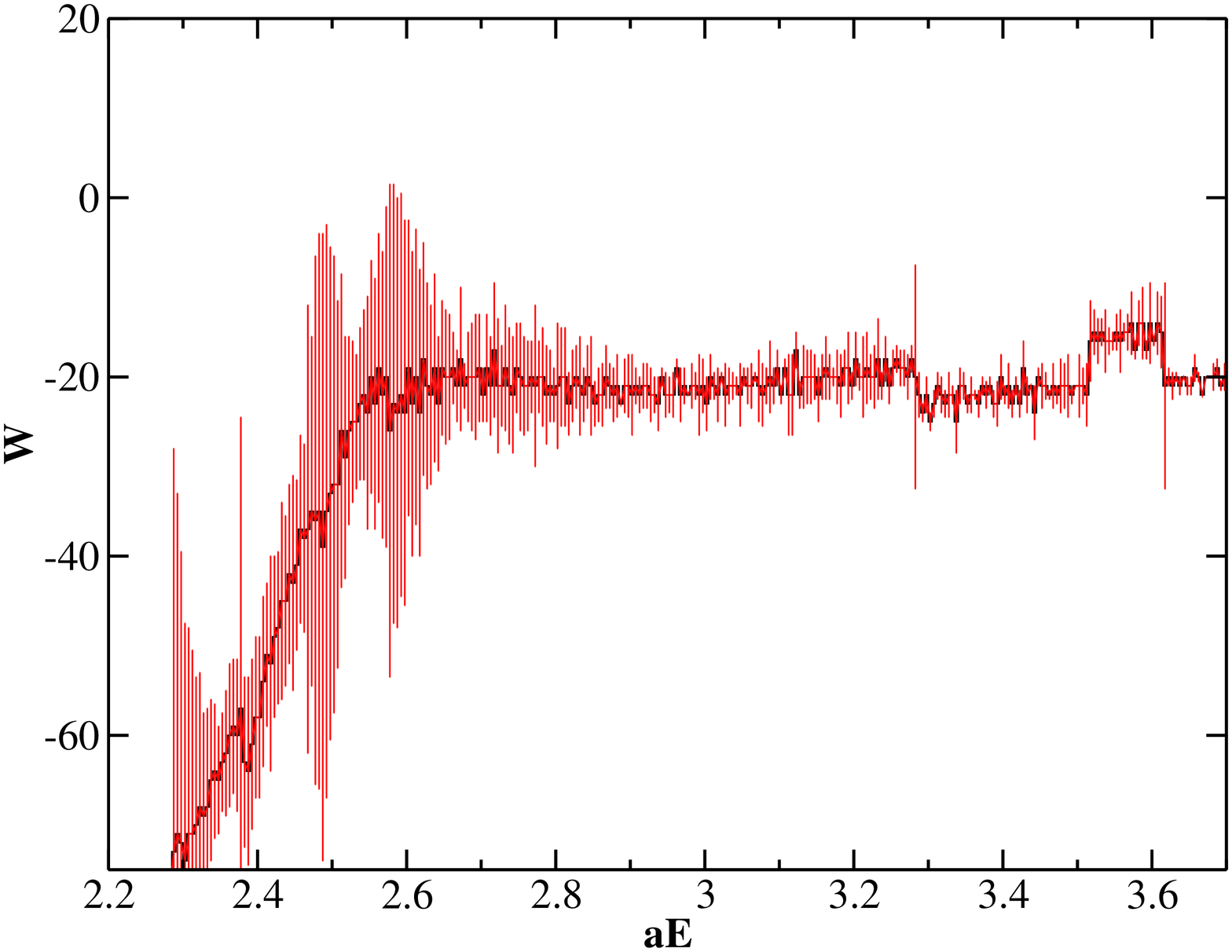}
\caption{(Top)  The probability distribution considering 
the correct background.  (Bottom)  Probability distribution 
$\tilde{W}$ using the correct background and excluding the two levels 
that are without a corresponding background. Simulation parameters:
 $\nu=1.0$, $\lambda=200.0$, $am_{\pi,0}=0.86$.}
\label{L10b}
\end{figure}

We repeat the procedure described above. Taking care of the
correct subtraction of the background we get the histogram of 
Figure~\ref{L8a} (Bottom). Clearly we see two discontinuities, related to the 
two levels in the interacting theory that appear without a corresponding 
background: one at $E \approx 1.95$ is due to the intersection at $L=6a$ and 
the other one at $E \approx 2.00$ which is due to the intersection at $L=20a$.
When we exclude the two levels which have no corresponding background signal,
we get the histogram of Figure~\ref{L8c} (Top). 
In this case again we can clearly see a Breit-Wigner shape and we can fit 
these data, as shown in Figure~\ref{L8c} (Bottom), obtaining the 
following parameters: 
$aM_\sigma=2.01(2)$, $a\Gamma_\sigma=0.35(10)$.
Finally, a third series of simulations was performed with parameters
$\nu=1.0$, $\lambda=200.0$, $am_{\pi,0}=0.86$. They have been tuned to
have the intersection between the $\sigma$ energy level and
$(2,0,0)$ two-particle energy level around $L=10a$. Because in this case
we are considering pions with higher momentum, we expect the width of the 
resonance to be larger than the previous cases, for reasons discussed at the 
end of Sec~\ref{themodel}. For this analysis, we take into account 13 levels 
to describe the shape of the resonance better.
In Figure~\ref{L10a} the spectrum for $6 \leq L/a \leq 15$ is plotted.
The onset value for the plateaux is $t_0 = 1$ and the relative error 
varies in the range 0.15\% - 0.4\%.
The measured mass for the pion is $am_\pi=0.938(3)$.
In Figure~\ref{L10b} (Top) as in the previous cases we show the probability
distribution taking in account all levels. We can see that a possible 
peak is present around a value of the mass $am \approx 2.8$.
Unfortunately, as seen in Figure~\ref{L10b} (Bottom), when we 
exclude the two levels the probability
distribution plot is flat and no Breit-Wigner shape emerges.
It is clear that in this case, the only way to determine the parameters
of the resonance is to considerably increase the number of measurements
and consequently to decrease the relative errors in the spectrum determination.

\subsection{L\"uscher's method results}\label{numres}
As outlined in Sec~\ref{luschmethod}, L\"uscher's method provides a way 
to relate information on the two-particle spectrum in the elastic region to 
the scattering phase shift. As the scattering phase shift depends on momentum
the first step is to convert the energy spectra data into momentum spectra data.

The relation between the energy and the momentum is given by the dispersion 
relations; however there is the choice of using the lattice dispersion 
relations or the continuum dispersion relations. Naturally the lattice 
dispersion relations are seen to better represent the data, but it is 
interesting to observe what occurs when the continuum dispersion relations are 
used.
\begin{figure}[htb]
  \includegraphics[width=0.35\textwidth,angle=-90]{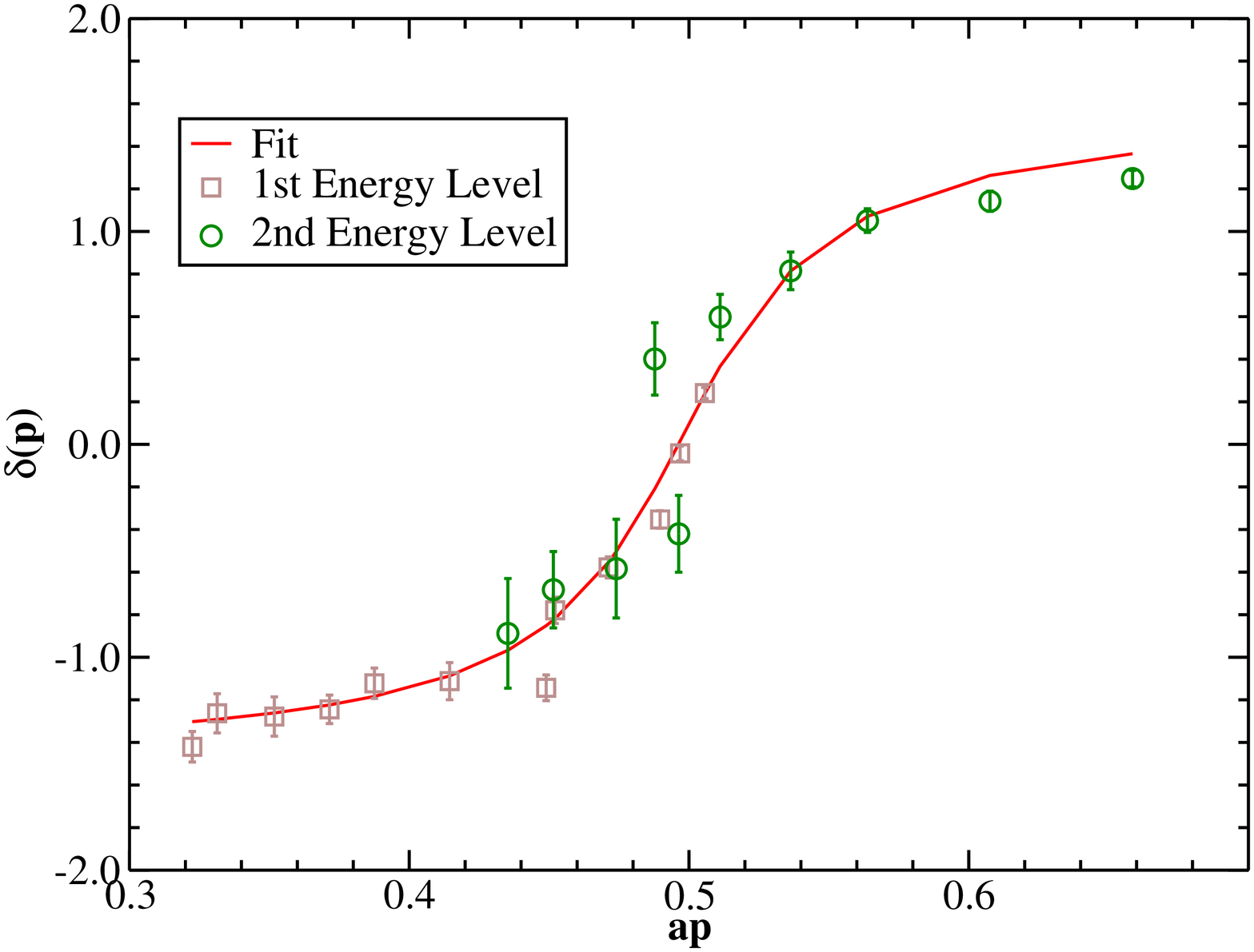}
\hspace{2mm}
  \includegraphics[width=0.35\textwidth,angle=-90]{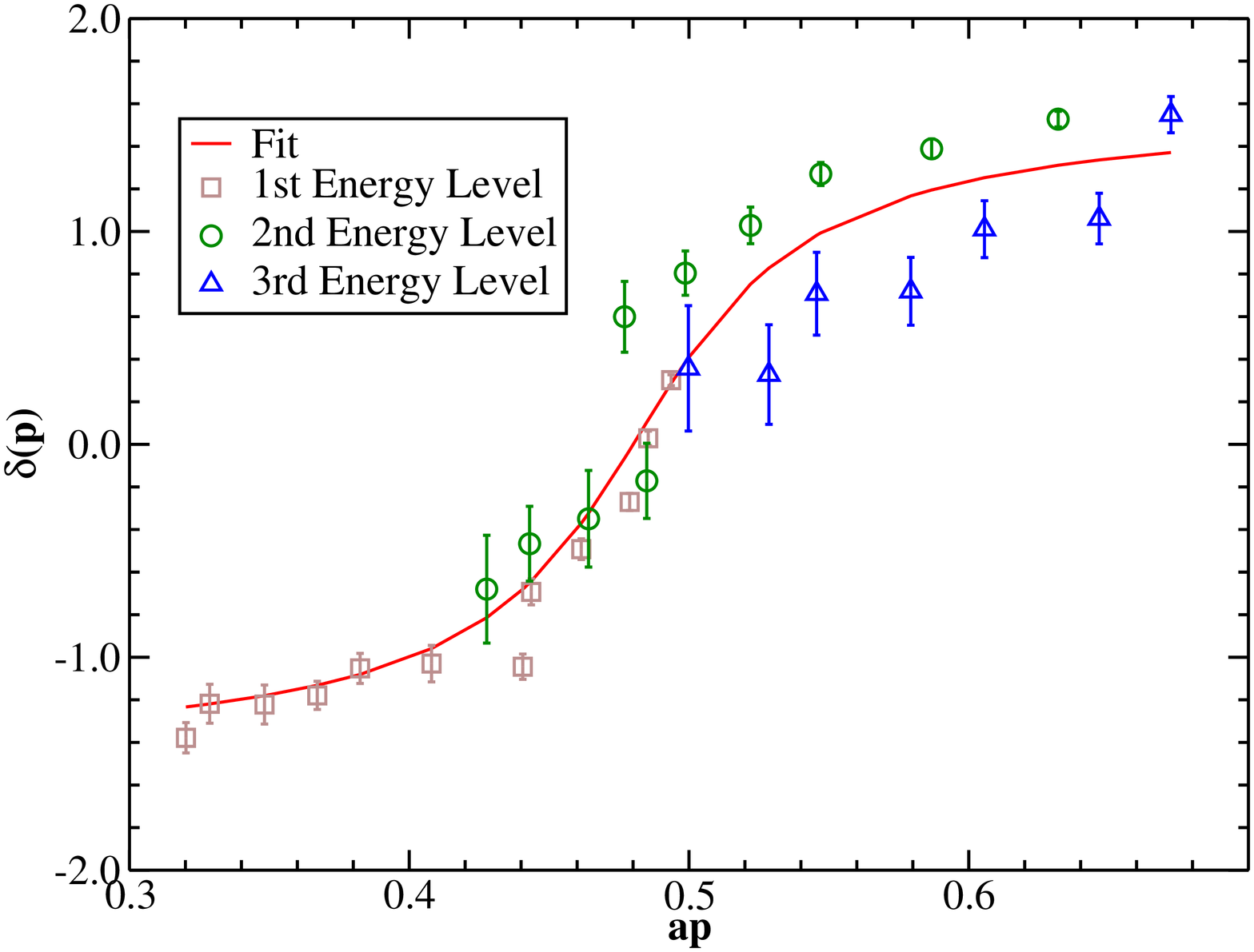}
\caption{(Top) $\delta(p)$ using Lattice dispersion relations at:
$\nu=1.0$, $\lambda=1.4$,
$am_{\pi,0}=0.36$. (Bottom) Same parameters, but with continuum dispersion 
relations. Both done with our $\phi(\kappa)$ approximation.}
\label{Dispersion}
\end{figure}
When the momenta spectrum $p_n(L)$ has been obtained through the 
dispersion relations it is necessary to have some knowledge of the 
function $\phi(\kappa)$ appearing in Eq.~\ref{lusherformu} in order to 
translate to the scattering phase shift $\delta(p)$. 
In some works, $\phi(\kappa) \approx \pi \kappa^{2}$ is taken as a good 
approximation, but it is possible that for low values of $\kappa$ this 
will not be sufficiently accurate.
For more accurate results one should numerically evaluate $\phi(\kappa)$. In essence
this amounts to a numerical evaluation of $\mathcal{Z}_{00}(r;q^{2})$. However, in the expression 
Eq.~\ref{ZETA} for $\mathcal{Z}_{00}(r;q^{2})$ the value of $r = 1$,
used in the application of L\"uscher's method, is outside
the domain of convergence.

Fortunately there is an integral representation of $\mathcal{Z}_{00}(1;q^{2})$
(Appendix C of Ref~\cite{Luscher:1990ux}) which analytically continues 
to the point $r = 1$. The expression is also amenable to numerical 
evaluation. Using the values obtained from this evaluation of 
$\mathcal{Z}_{00}(1;q^{2})$, we performed a fit of $\phi(\kappa)$ to 
obtain as our approximation in the range $\kappa \in \left[0.1,1.5\right]$
\begin{eqnarray}
\phi(\kappa) 
\approx 
(-0.09937)\kappa^{8} + (0.47809)\kappa^{6} + \nonumber \\
(-0.62064)\kappa^{4} + (3.38974)\kappa^{2} \ .
\label{eq:phi_fit}
\end{eqnarray}
Note that it can be shown from its definition that $\phi(\kappa)$ has a Taylor
expansion consisting of only even powers of $\kappa$. The error of using this 
approximation in place of the true values, within the given range, 
is significantly less than other errors and can be neglected at later stages 
of the analysis. This approximation is used here simply to demonstrate the deviation
of the function from $\phi(\kappa) \approx \pi \kappa^{2}$ and how
this can affect the results.\\ 
We can now use Eq.~\ref{lusherformu} to map our energy spectrum data to
$\delta(p)$. The choice of dispersion relations and approximation of 
$\phi(\kappa)$ could possibly change the results significantly so all 
four choices are considered.

\begin{table}[htb]
\begin{center}
\begin{tabular}{cc@{\hspace{2em}}cc@{\hspace{2em}}cc}
\hline
      &           & \multicolumn{2}{c}{$\phi(\kappa)$} & 
     \multicolumn{2}{c}{$\pi\kappa^2$} \\
$\nu$ & $\lambda$ &   $aM_\sigma$ & $a\Gamma_\sigma$ & 
       $aM_\sigma$ & $a\Gamma_\sigma$ \\
\hline
1.0   & 1.4       & 1.32(8) & 0.117(9) & 1.4(1) & 0.16(5) \\
1.0   &   4       & 2.1(4) & 0.39(4) & 2.2(4) & 0.42(5) \\
1.0   & 200       & 3(1) & 1.2(7) & 3(1) & 2(2) \\
\hline
\end{tabular}
\end{center}
\caption{Resonance mass and decay width using two different approximations for $\phi(\kappa)$, with continuum
dispersion relations.}
\label{table2}
\end{table}
Firstly, for the choice of dispersion relations, Figure~\ref{Dispersion} shows
that the lattice relation brings the energy levels close to a single arctangent 
profile whereas the continuum relations give a much more scatter. Also notice 
that the third energy level is not mapped to the elastic region with the 
lattice dispersion relations.  The lattice dispersion relations also have 
smaller errors.
\begin{table}[htb]
\begin{center}
\begin{tabular}{cc@{\hspace{2em}}cc@{\hspace{2em}}cc}
\hline
      &           & \multicolumn{2}{c}{$\phi(\kappa)\quad$} & 
     \multicolumn{2}{c}{$\pi\kappa^2$} \\
$\nu$ & $\lambda$ &   $aM_\sigma$ & $a\Gamma_\sigma$ & 
       $aM_\sigma$ & $a\Gamma_\sigma$ \\
\hline
1.0   & 1.4       & 1.35(2) & 0.115(8) & 1.36(4) & 0.17(2) \\
1.0   &   4       & 2.03(2) & 0.35(2) & 2.2(2) & 0.42(5) \\
1.0   & 200       & 3.1(7) & 1.2(5) & 3(1) & 2(1) \\
\hline
\end{tabular}
\end{center}
\caption{Resonance mass and decay width determinations using two different 
approximations for $\phi(\kappa)$ and lattice free dispersion relations.}
\label{table3}
\end{table}

After fitting, the results for the resonance mass and decay width in the 
two approximations using continuum dispersion relations are shown in 
Table~\ref{table2} and the lattice dispersion relations in 
Table~\ref{table3}.

The choice of approximation for $\phi(\kappa)$ can be seen to most strongly 
affect the errors and values of the decay width.
A possible reason for this is that different approximations 
of $\phi(\kappa)$ will change the slope of the scattering phase shift, 
which is directly related to the the decay width of the resonance. 
So it would appear that the lattice dispersion 
relations should be used for a clear arctangent profile and a good approximation
to $\phi(\kappa)$, such as Eq.~\ref{eq:phi_fit}, so that the slope remains 
undistorted to give accurate information on the decay width.

\begin{figure}[htb]
  \includegraphics[width=0.35\textwidth,angle=-90]{./figure/phaseshift/Lattice_Correct.ps}
  \includegraphics[width=0.35\textwidth,angle=-90]{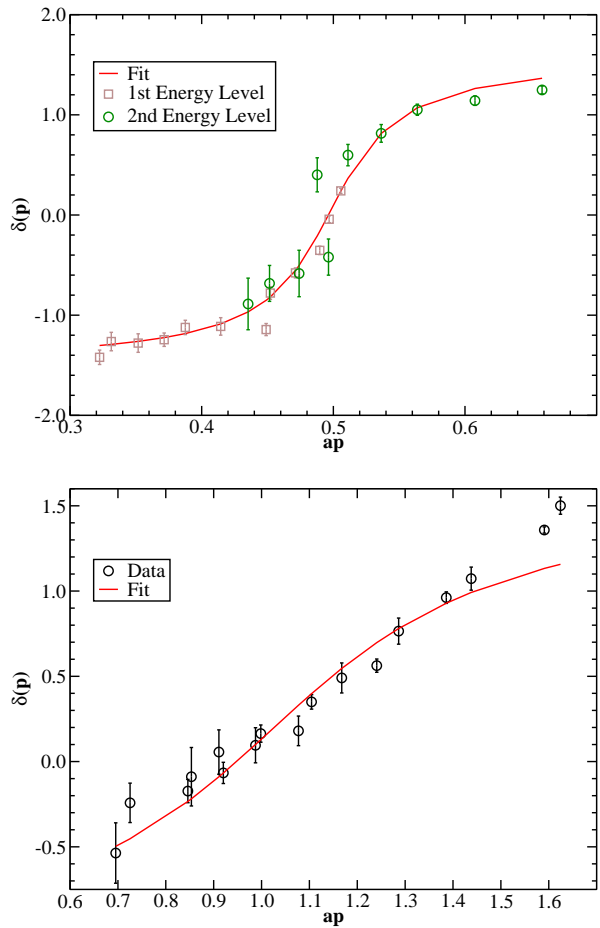}
\caption{Phase shift for narrow ($\nu=1.0$, $\lambda=1.4$, $am_{\pi,0}=0.36$) 
and broad ($\nu=1.0$, $\lambda=200.0$, $am_{\pi,0}=0.86$) resonances.}
\label{Broad}
\end{figure}

It can be seen that the errors increase as the resonance gets broader. 
Similar to
the histogram method this is related to the distinctive profile of the resonance
being washed out. In the histogram method the distinctive Breit-Wigner 
form flattened out into a flat profile, here the typical arctangent profile 
of the phase shift becomes a straight line. In this case the resonance width 
can be changed within a wide margin without affecting the profile of the 
phase shift, hence the greater errors. Figure~\ref{Broad}
shows a comparsion between the broadest case and the narrowest case.

\subsection{Comparison}

Results from a comparison between L\"uscher's method and the histogram method 
are shown in Table~\ref{table4}.
\begin{table}[htb]
\begin{center}
\begin{tabular}{cc@{\hspace{2em}}cc@{\hspace{2em}}cc}
\hline
 & & \multicolumn{2}{c}{L\"uscher\hspace{2em}{ }} & \multicolumn{2}{c}{histogram}\\
$\nu$ & $\lambda$ & $aM_\sigma$ & $a\Gamma_\sigma$ & 
   $aM_\sigma$ & $a\Gamma_\sigma$ \\
\hline
1.0 & 1.4 & 1.35(2) & 0.115(8) & 1.33(5) & 0.10(5) \\
1.0 & 4   & 2.03(2) & 0.35(2)  & 2.01(2) & 0.35(10) \\
1.0 & 200 &  3.1(7) & 1.2(5)   &   ---   &   --- \\
\hline
\end{tabular}
\end{center}
\caption{A comparison between the L\"{u}scher and the histogram method. For
the very broad resonance, no determination of the resonance parameters was obtained.}
\label{table4}
\end{table}
L\"uscher's method gives smaller errors than the histogram 
method, but the results are broadly consistent. L\"uscher's method 
manages to provide some estimate on the width of the resonance in the 
broad case. The broad resonance becomes a problem for the histogram 
method because there is no obvious peak to indicate the resonance mass 
and hence no width of that peak to determine the decay width. 
One would need very precise data in order to avoid a washing out of
the structure of the histogram. 
L\"uscher's method also becomes more difficult to apply in the case 
of broad resonances. Here, the profile 
of $\delta(p)$ is quite flat, hence a large range of parameters will 
be capable of fitting to the profile. Again an accurate determination 
of the energy levels is required to determine the profile precisely 
enough so that this is prevented.
Considering the amount of work necessary until one can use the 
histogram method (as detailed above), L\"uscher's method is considerably 
easier to apply, provided one has a good approximation of $\phi(\kappa)$. 
However, the histogram method can be used as a visual tool for 
spotting the resonance.

One restriction of L\"uscher's formula is that it only applies in the 
elastic region. It is possible that the histogram 
method will provide a means of determining the presence of a resonance 
in the inelastic region. Certainly, a histogram can be constructed in the 
inelastic region; the only difficulty is that with the inapplicability 
of L\"uscher's formula it is unclear that the parameters of this 
histogram will have any relation to those of the resonance.

\subsection{Inelastic scattering}

We want to discuss now what happens when we tune the resonance parameters
to have a mass resonance greater then four times the pion mass, i.e. 
greater then the elastic threshold.

We run a series of simulations with parameters
$\nu=1.05$, $\lambda=0.85$, $am_{\pi,0}=0.17$.
They have been tuned to have the intersection between the $\sigma$ 
energy level and $(1,0,0)$ two-particle energy level around $L=11a$.

\begin{figure}[htb]
\begin{center}
\hspace{-2mm}
  \includegraphics[width=0.36\textwidth,angle=-90]{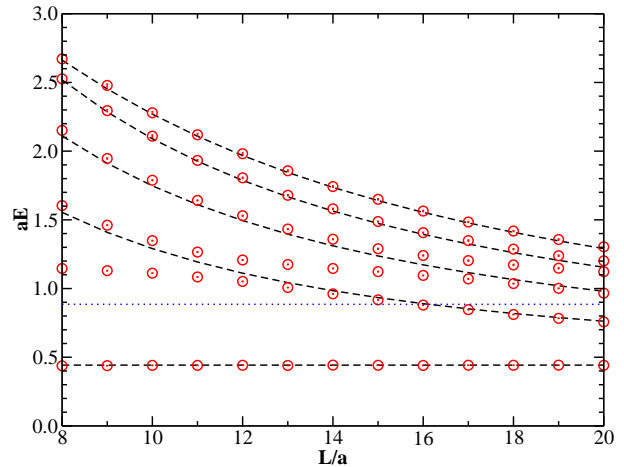}
\caption{Spectrum of the theory for different values of the
volume for the following simulation parameters, describing an inelastic 
scattering: $\nu=1.05$, $\lambda=0.85$, $am_{\pi,0}=0.17$. 
The horizontal blue dotted line shows the elastic threshold.}
\label{anelspectr}
\end{center}
\end{figure}

The physical mass for the pion turns out to be $am_\pi=0.2213(5)$.
In Figure~\ref{anelspectr} we plot the spectrum for $8 \leq L/a \leq 20$
for the first six levels; in this case the onset value for the plateaux 
is $t_0=2$ and the relative error varies in the range 0.08\% - 0.4\%.

For L\"uscher's method the results are nonsensical, as would be expected 
since the method can only be demonstrated in the elastic region due to 
the restricitions of the Bethe-Salpeter Kernel and more fundamentally the 
fact that the formula is first derived in quantum mechanics.

\begin{figure}
\begin{center}
  \includegraphics[width=0.35\textwidth,angle=-90]{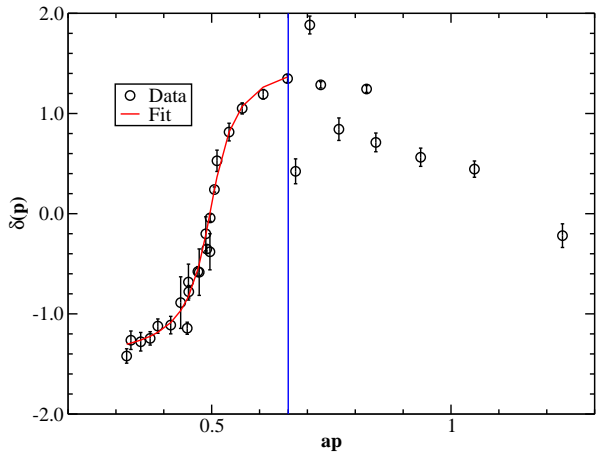}
\caption{Inelastic data with L\"uscher's formula. For the case of $\nu = 1.0$, $\lambda = 1.4$, $am_{\pi,0} = 0.36$. 
(Onset of inelastic region marked).}
\label{Inelastic}
\end{center}
\end{figure}

Figure~\ref{Inelastic} shows an example of applying the method beyond 
the inelastic threshold for $\nu=1.0$ and $\lambda = 1.4$. It can be seen 
that the profile does not fit what would expected of the
scattering phase shift and in fact the second point after the threshold, 
being above $\pi/2$, could even break unitarity.
Fortunately for these values of $\nu$ and $\lambda$ the resonance is not above the treshold. For
$\nu=1.05$, $\lambda=0.85$, where the resonance is above threshold, the problems mentioned above
make the results uninterpretable.
It is worth noting that in Figure~\ref{anelspectr} we do not have any 
hints of the expected $4 \pi$ level; an explicit implementation of an 
interpolator should therefore be necessary.

\begin{figure}[htb]
\hspace{-2mm}
  \includegraphics[width=0.36\textwidth,angle=-90]{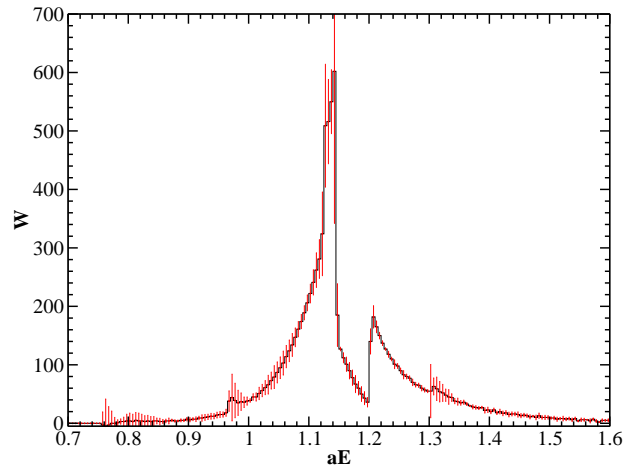}
\hspace{2mm}
  \includegraphics[width=0.36\textwidth,angle=-90]{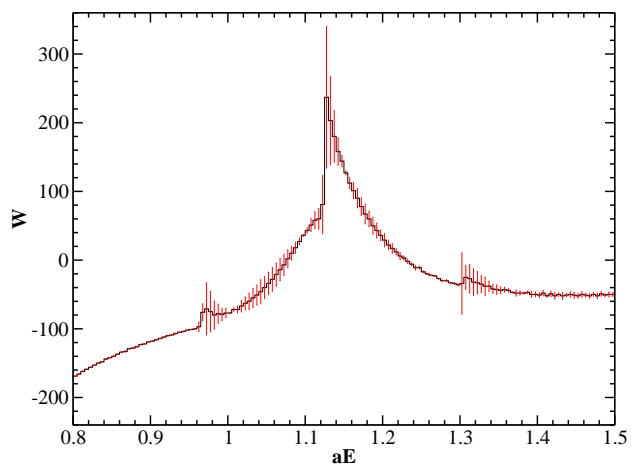}
\caption{(Top) The probability distribution considering 
the correct background.  (Bottom)  Probability distribution 
$\tilde{W}$ using the correct background and excluding the two levels 
that are without a corresponding background. Simulation parameters: 
$\nu=1.05$, $\lambda=0.85$, $am_{\pi,0}=0.17$.}
\label{anelhist1}
\end{figure}

In Figure~\ref{anelhist1}~(Top) the probability distribution is shown
considering the correct background and all levels. 
The distribution without the two levels, characterized by the absence
of their own background, is shown in Figure~\ref{anelhist1}~(Bottom).
In this case a bad and unexpected result is obtained: a jump
around $m \approx 1.13$ is present; this is a further
proof of how laborious this method is.

The background was subtracted following the same procedure as before but 
around $L=20a$ for the energy levels around $(1,1,1)$ a new problem arises.
We have already seen that the only way to avoid a jump in the 
histogram $\tilde{W}$ 
is to make the correct correspondence at the two extremities of the volume 
interval ($L=8a$ and $L=20a$ in this case) between the energy levels of the 
interacting theory and the corresponding background. They should coincide
or at least be parallel (after the lengthening of the free spectrum lines). 
Figure~\ref{L12spectrcorr}~(Top) demonstrates that this is exactly what 
happened in the previous cases; 
this characteristic is not present in this case. The two lines are not 
parallel because for $L=20a$ the effect of the interaction is too strong.
Note that this problem is not related to the inelastic regime, 
but could be present in the previous cases as well; 
it is only by chance that this did not happen.
The only way to avoid this new problem is to consider a different 
volume range. In particular we have verified that in this case a better 
choice is $8 \le L/a \le 18$.
\begin{figure}[htb]
\hspace{-2mm}
  \includegraphics[width=0.36\textwidth,angle=-90]{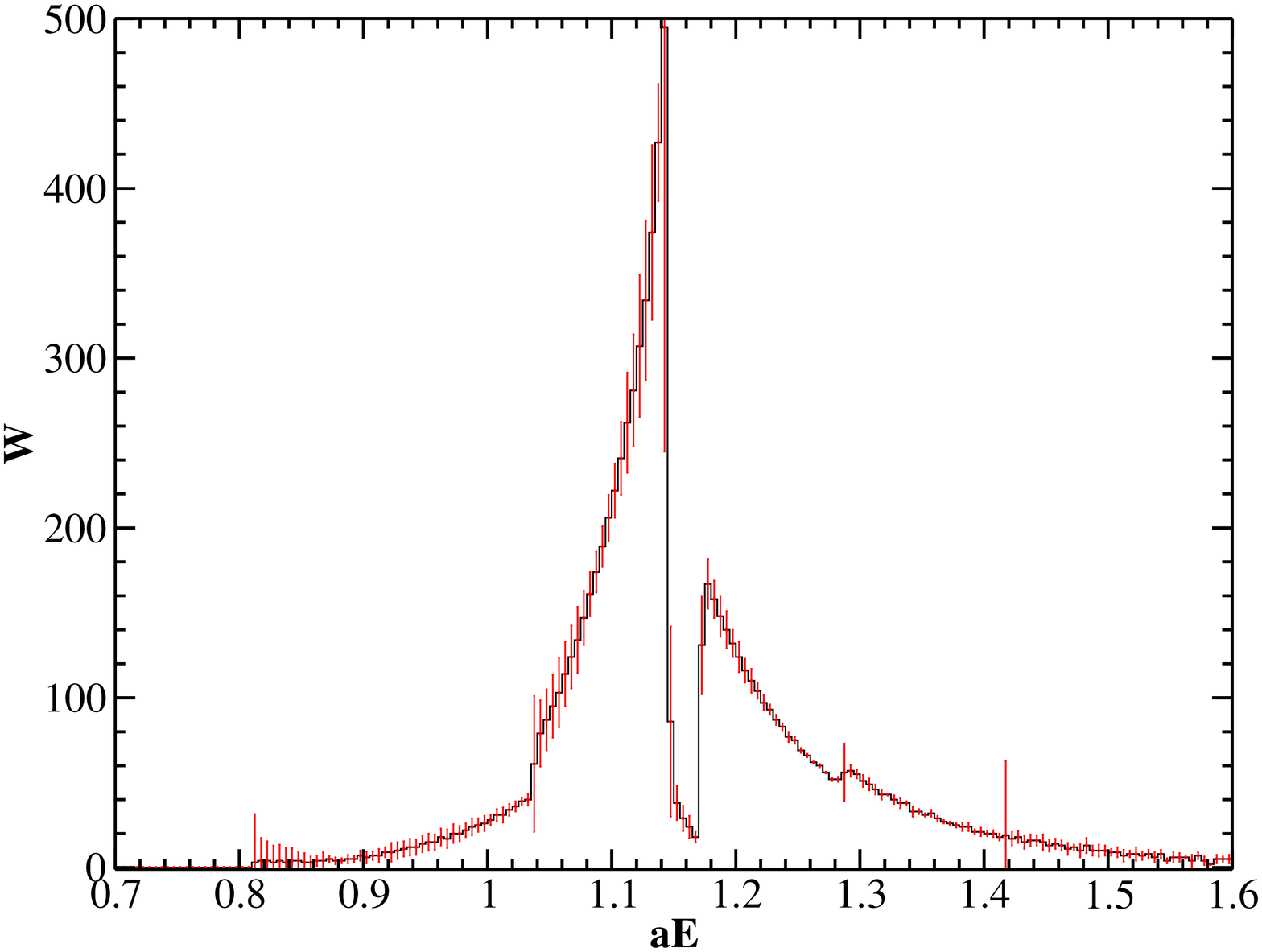}
\hspace{2mm}
  \includegraphics[width=0.36\textwidth,angle=-90]{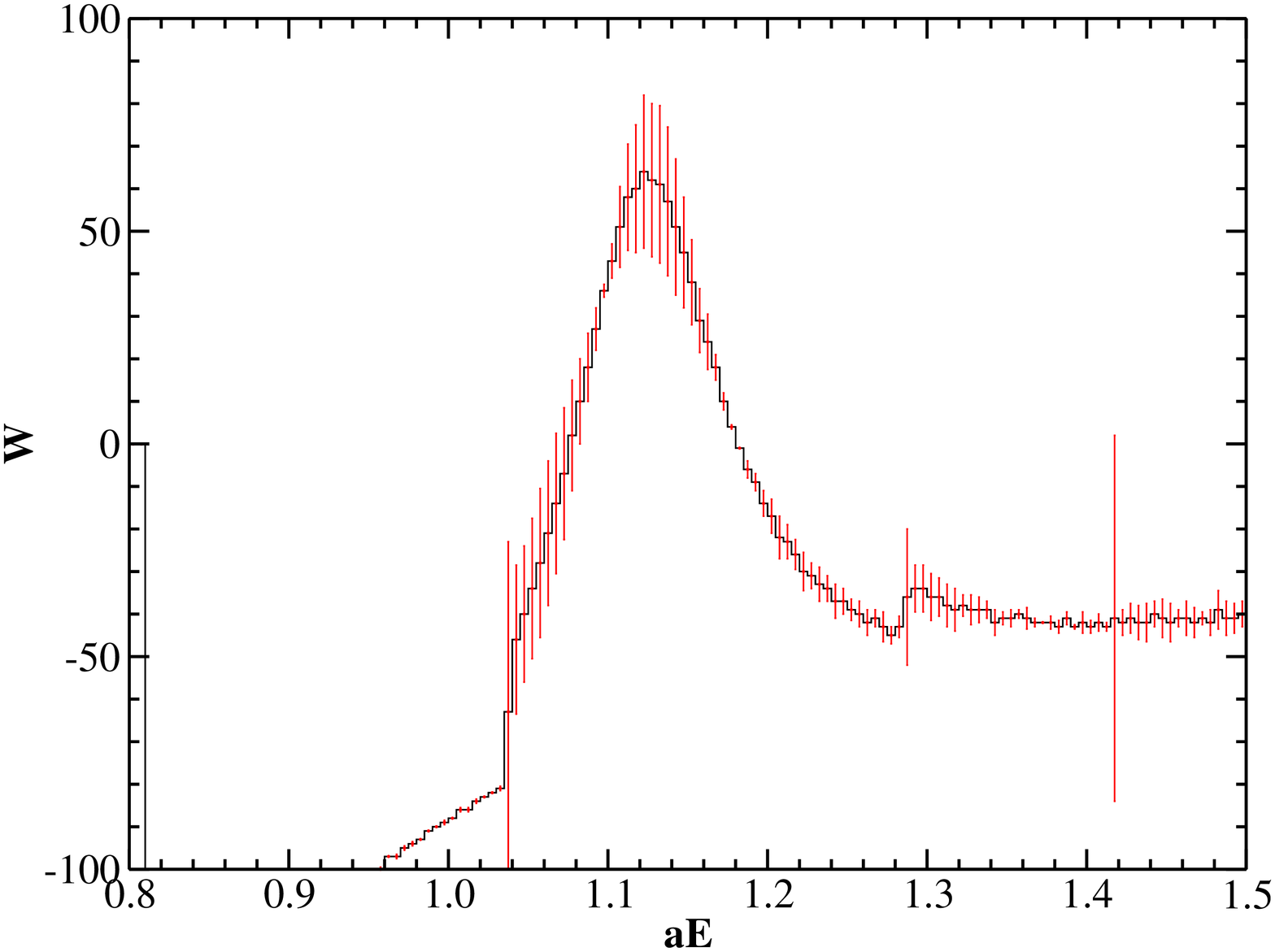}
\caption{Like Figure~\ref{anelhist1} but considering a volume range 
$8\le L/a \le 18$ in Figure~\ref{anelspectr}. Simulation parameters: 
$\nu=1.05$, $\lambda=0.85$, $am_{\pi,0}=0.17$.}
\label{anelhist2}
\end{figure}
Using this new interval we can determine the probability distribution
shown in Figure~\ref{anelhist2}~(Top); excluding the two levels as before
we get the result of Figure~\ref{anelhist2}~(Bottom). Finally, we can see a
Breit-Wigner shape that we can fit as shown in 
Figure~\ref{anelhistfit} obtaining the following parameters: 
$aM_\sigma=1.11(3)$, $a\Gamma_\sigma=0.11(3)$.
\begin{figure}[htb]
\hspace{-2mm}
  \includegraphics[width=0.36\textwidth,angle=-90]{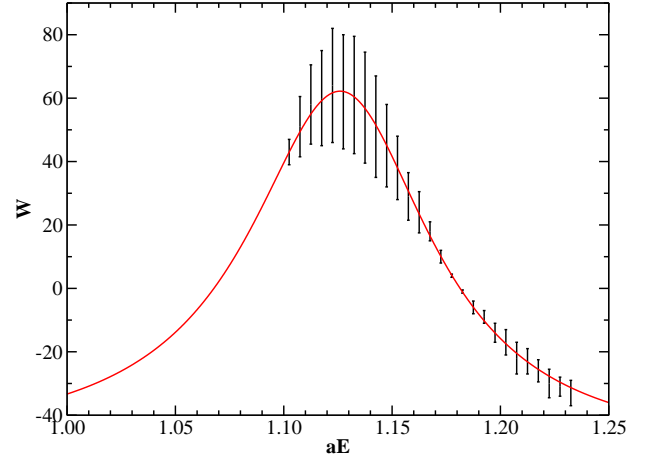}
\caption{Data from Figure~\ref{anelhist2} (Bottom) that we fit to 
determine the resonance parameters with the final curve fitting. 
Simulation parameters: $\nu=1.05$, $\lambda=0.85$, $am_{\pi,0}=0.17$.}
\label{anelhistfit}
\end{figure}
It is clear therefore that in the inelastic regime we can also apply
exactly the same procedure as was developed for the elastic case and finally
we can determine in this case the resonance parameters also.
Unfortunately, in contrast with the elastic regime there is no theoretical
support in this case. Therefore, even if we can determine the
parameters for the Breit-Wigner shape of the probability distribution
hystogram, there is no reason to link these numbers with the resonance
parameters.
\subsection{Correlator method}
\begin{figure}[htb]
\hspace{-2mm}
\includegraphics[scale=0.32,angle=-90]{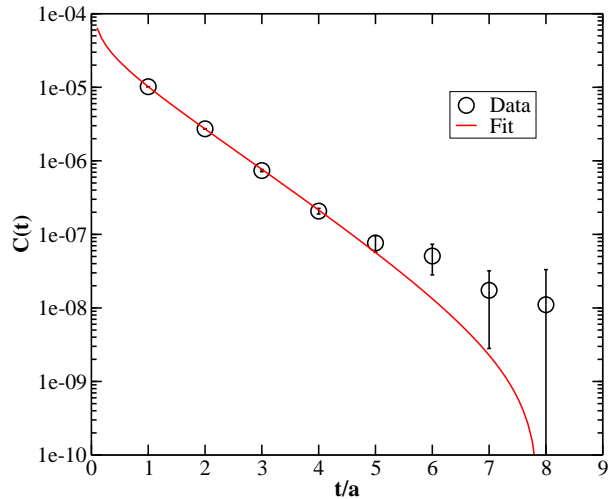}
\caption{Fit to sigma correlator; parameters: 
$\nu=1.0$, $\lambda=1.4$, $am_{\pi}=0.36$.}
\label{Corrmethfit}
\end{figure}

We have also performed a preliminary investigation of a third method 
described in Ref~\cite{Meissner:2010rq}. This method attempts to extract 
resonance parameters via fitting the correlator to some asymptotic form
at small times. This avoids the Maiani-Testa theorem, as the theorem only 
restricts access to scattering information via the $n$-point functions 
with $n\geq3$. Also, in using the correlator we are treating
resonances on the same footing as stable states. The form of the correlator 
that we fit to is:
\begin{eqnarray}
D(t)&=&\mbox{e}^{-\omega_{\sf min}t}\,\biggl\{
c_0 F^{(0)}(t,E_R)+
\nonumber \\
& & 
c_1 F^{(1)}(t,E_R)
+\sum_{k=0}^\infty\frac{x_k}{t^{l+k+3/2}}\biggr\} \ , 
\label{eq:representation}
\end{eqnarray}
$\omega_{min}$ being the value of the multiparticle threshold, which
in this work is $\omega_{min} = 2m_{\pi}$; $E_R$ is the location of
the pole associated with the resonance relative to the multiparticle
threshold, namely $E_{R} = \left(M_{\sigma} - \omega_{min}\right) - i\frac{\Gamma}{2}$, with
$M_{\sigma}$ and $\Gamma$ the mass and width of the resonance respectively. We label
the real part of $E_{R}$ as $E_{0}$ in what follows, $E_{0} = \left(M_{\sigma} - \omega_{min}\right)$.\\
The $F^{(i)}(t,E_R)$ functions have the following definition:
\begin{eqnarray}
F^{(0)}(t,E_R)&=&-\frac{2}{\Gamma}\, \mbox{Im}\,\chi(t,E_R)
\nonumber \\[2mm]
F^{(1)}(t, E_R)&=&\mbox{Re}\,\chi(t, E_R)-\frac{2 E_{0}}{\Gamma}\,\mbox{Im}\,
\chi(t,E_R) \ .
\end{eqnarray}
The function $\chi(t, E_R)$ is calculated via the expression:
\begin{eqnarray}
\chi(t,E_R)&=&
-\pi\,\sqrt{-E_R}\, \mbox{e}^{-E_R t}+ 
\nonumber \\
& & \sqrt{\frac{\pi}{t}}\,\biggl\{1+\sum _{i=0}^{\infty}\, 
\frac{(-2 E_R t)^{i+1}}{(2i+1)!!}\biggr\}\, .
\label{chiexp}
\end{eqnarray}
A resonance is to be found as a pole on the second (unphysical) Riemann 
sheet of the Kallen-Lehmann spectral function. However since the branch 
cut that gives rise to this second Riemann sheet dissolves into a series
of poles in finite volume, it is not obvious how the resonance can have 
an effect on the correlator. However if, in infinite volume, the resonance 
is very narrow then it is close enough to the branch cut for it to have 
an effect on the first (physical) Riemann sheet and so its influence will 
show up in the infinite volume correlator. The finite volume correlator 
converges to the infinite volume one rapidly at large volumes and so the 
influence of the resonance shows up in the finite volume correlators we 
observe on the lattice. We should then be able to apply 
Eq.~\ref{eq:representation} at large volumes.
In Eq.~\ref{eq:representation} the $x_k$ represent the non-resonant 
scattering, which in our model we expect to be small. We chose the 
value $k = 2$, as smaller values were found to give poor results. 
We then fitted the sigma correlator for the $L = 19$ lattice for the $\nu=1.0$, 
$\lambda = 1.4$, $am_{\pi} = 0.36$ parameters
and obtained the following results (the fit is shown in Fig.~\ref{Corrmethfit}):
\begin{eqnarray}
aM_{\sigma} &=& 1.32(5)  \ , \nonumber\\
a\Gamma_{\sigma} &=& 0.107(7) \ , \nonumber\\
c_0 &=& -0.00122(4)  \ , \nonumber\\
c_1 &=& 0.00023(8)\ , \nonumber\\
x_0 &=& 0.078(1) \ , \nonumber\\
x_1 &=& 0.158(5) \ . \nonumber
\end{eqnarray}
The fit, which was done in the window $t \in [1,8]$, has a chi-squared per 
degree of freedom of $\chi^{2}/\nu=0.8362$. The resonance
width $\Gamma$ appeared to quite sensitive to the fit window if values of
$t$ greater than $10$ were taken. However this is possibly not a surprising result
as the form for the correlator Eq.~\ref{eq:representation} is derived for small times.\\
It should be noted that this method obtained these results via a fit to 
the correlator in a single, large volume. Of course the method also 
introduces new fitting parameters, $c_0, c_1, x_0$ and $x_1$, which 
make the fit less discriminating. The method also appears to be restricted to 
narrow resonances, attempts to apply the method to the broader resonance data 
of this work were not succesful.\\
The results are however consistent with the two preceeding methods.
Only a preliminary investigation of this method was made, in particular a more 
precise estimate of the errors via the Bayesian analysis suggested in 
Ref~\cite{Meissner:2010rq} might improve the situation.

\section{Conclusions and outlook}
\label{secconclusion}
The investigation of the two methods studied in this work has elucidated 
their relative strengths and weaknesses when they are applied to data from 
a Monte Carlo study, which have finite statistical precision. 
L\"uscher's method requires an estimation on the functional form of the 
scattering phase shift, in our case we used the ansatz of the 
Breit-Wigner form associated to an isolated resonance. Once these two
requirements are met the method is relatively straight forward to apply. 
The main disadvantage is the increasing errors as the resonance 
becomes broader and the clear restriction to studying elastic scattering.

The histogram method, since it has the form of a Breit-Wigner peak, 
provides a distinctive visual check of the presence of resonances. However the 
method of constructing a histogram from Monte Carlo data with a limited range 
of volumes is not as straightforward as applying L\"uscher's technique and one 
also finds increasing errors for broader resonances. For very broad 
resonances, the method misses the state entirely. 

L\"uscher's method is then the stronger of the two based on our experience 
here due to its ease of application. For narrow resonances however, results 
from both techniques appear to be complimentary and have similar statistical 
precision. We briefly investigated a third method which makes more direct 
use of the time-dependence of correlator data and would treat resonances 
similarly to stable states, but much remains to be done to show it is useful
for the analysis of Monte Carlo data. 

The major drawback for all methods is that they are restricted to 
the elastic region. Studing the inelastic region is of crucial importance to 
learning more detail about the resonances that emerge from QCD. What is clear
is that any more advanced method that has potential in that region will need
to be able to deal with statistical uncertainty in a robust way without the
need for delicate fine-tuning.

\section*{Acknowledgements}
This work is supported by Science Foundation Ireland under
research grant 07/RFP/PHYF168. We are grateful for the continuing support of
the Trinity Centre for High-Performance Computing, where the numerical
simulations presented here were carried out.

\bibliography{scatter}

\end{document}